\documentclass[%prl,
 aps,
 ,twocolumn
]{revtex4}
\usepackage{graphicx}
\usepackage{amsmath}
\usepackage{amssymb}
\usepackage{mathrsfs}
\usepackage{dsfont}
\usepackage{appendix}
\usepackage{tikz}
\usepackage{pgflibraryarrows}
\usepackage{pgflibrarysnakes}
\usetikzlibrary{decorations.pathmorphing}
\usepgflibrary{arrows.meta}

\usepackage{subfigure}

\usepackage{braket}

\begin{document}

\title{Universal Transformation of Displacement Operators and its Application to Homodyne Tomography In Differing Relativistic Reference Frames}
%\title{Accelerated time-delay hides information through correlations in vacuum noise}

%\author{Daiqin Su and T.C.Ralph}
\author{Sho Onoe}
\email{sho.onoe@uqconnect.edu.au}
\author{Timothy.~C.~Ralph}\email{ralph@physics.uq.edu.au}
\affiliation{Centre for Quantum Computation and Communication Technology, School of Mathematics and Physics, The University of Queensland, St. Lucia, Queensland, 4072, Australia}

\date{\today}

\begin{abstract}
{In this paper, we study how a displacement of a quantum system appears under a change of relativistic reference frame. We introduce a generic method in which a displacement operator in one reference frame can be transformed into another reference frame. It is found that, when moving between non-inertial reference frames there can be distortions of phase information, modal structure and amplitude. We analyse how these effects affect traditional homodyne detection techniques. We then develop an in principle homodyne detection scheme which is robust to these effect, called the ideal homodyne detection scheme. We then numerically compare traditional homodyne detection with this in principle method and illustrate regimes when the traditional homodyne detection schemes fail to extract full quantum information.}
\end{abstract}
\maketitle
%\vspace{10 mm}
\section{Introduction}
The advancement of quantum information (QI) over the past three decades has led to informational resources, processes and storage \cite{WildeQIT} beyond the classical limit. 
%One of the most remarkable result of QI may be the increased sensitivity to gravitational waves \cite{LIGO}. 
Broadly speaking, QI may be characterised as discrete variable (DV) or continous variable (CV). DV \cite{Nielson} is where information is encoded in discrete, finite degrees of freedom. Whilst CV entails encoding information in continuous degrees of freedom \cite{Braunstein2005, Weedbrook2012, Adesso2014}. %In terms of quantum field theory, fermionic fields can be considered as DV while bosonic fields can be considered as CV. 
%The increased sensitivity of gravitational waves is a feat of CV-QI metrology.
\\
\\
Amongst the CV-QI processes, there is a special class of states; Gaussian probability states. So-called because their quadrature distributions follow Gaussian statistics. Gaussian operations and measurements are those which preserve the Gaussianity of states. Gaussian states exhibit nice features which leads to various benefits to both theoreticians and experimentalist. For theorist, simple analytical tools are avalible and Gaussian states have special features which make analysis simple \cite{Weedbrook2012}.  For experimentalists, particularly in optics, all Gaussian operations can be reduced to local phase transitions, squeezers and beam splitters \cite{BMD1964, BraunsteinBMD}. Due to these properties, Gaussian QI has wide application to various fields \cite{Weedbrook2012}. These include quantum communication \cite{Briegel1998, Josephine2017, PatronThesis, QCMC}, quantum cryptography \cite{Ralph1999,Ralph2000}, quantum computation \cite{Lloyd1999}, quantum teleportation \cite{Ralph1999AOT}, quantum state and channel discrimination \cite{Childs2000}, and quantum metrology \cite{LIGO}. Of particular importance here is that change of relativistic reference frames preserve Gaussian states.
\\
\\
Measurement techniques are an essential feature of QI protocols. The key Gaussian CV-QI measurement technique is homodyne detection \cite{Welsch2009}. With the significant advancement in CV-QI over the past decades, we have started to consider quantum communication to reference frames which are non-inertial \cite{Onoe2018, Daiqin2017b, Hotta2015, Walker1985, Carlitz1987, Wilczek1992, Nicolaevici2003}. In these regimes, we observe surprising effects when we utilize traditional homodyne detection schemes.
\\
\\
A recent paper noted the notion of apparent decoherence \cite{Onoe2018, Daiqin2017b}, whereby a pure squeezed signal \cite{Daiqin2017b} or a time-delayed signal \cite{Onoe2018} created in the accelerated frame may seem decohered to an inertial observer. This decoherence effect was traced to the operational method in which the information was analysed; in particular, the self-homodyne detection scheme. The decoherence effect observed from an accelerated mirror was traditionally explained to be due to tracing out radiation that is reflected away from the observer \cite{Daiqin2017a,  Crispino2008}. However, it has been suggested that this decoherence may also be due to neglection of vacuum entanglement \cite{Onoe2018}. Self-homodyne detection fails to extract this entanglement. The existence of these vacuum entanglement was confirmed for a eternally perfect mirror moving along an exponentially accelerated trajectory \cite{Hotta2015}. These result suggest that the current homodyne detection scheme is incomplete, and we need to develop a complete homodyne detection scheme which accounts for changes to non-inertial reference frames.
\\
\\
In this paper we will develop a homodyne detection scheme which accounts for non-inertial changes in reference frames. We will do this by introducing a universal transformation of displacement operators. Our paper is set out in the following way. Section II, III and VI are dedicated to developing the technique we refer  to as ideal homodyne. Sections V compares the traditional homodyne methods to ideal homodyne detection. In section II, we introduce a method which allows universal transformation of displacement operators. We apply this technique to two examples. In section III, we review two well known homodyne detection schemes, and analyse how they can be applied to homodyne detection in differing reference frames. In section VI, by implementing the universal transformation of displacement operators, we develop a new detection scheme; ideal homodyne detections scheme. In section VA, we analytically compare the three detection scheme via analysing the QI of a coherent signal. In section VB, we produce numerical plots of interesting cases. We conclude and discuss future application of this technique in Section VI.
\section{Transformation of basis for displacement operators}
In this section, we consider a general method in which a transformation of basis can be conducted for displacement operators. We first consider an arbitrary normalised bosonic operator $\hat{O}$; $[\hat{O},\hat{O}^{\dag}]=1$. The displacement operator is then defined in the following way:
\begin{equation}
\begin{aligned}
\hat{D}_{\hat{O}}(\alpha) \equiv \exp(\hat{O}^{\dag}\alpha-\hat{O}\alpha^*)
\end{aligned}
\end{equation}
We now consider an arbitrary complete bosonic basis set $\hat{o}_{n,m}$, where $n$ is a discrete variable and $m$ is a continuous variable. Such a basis set satisfies the following property:
\begin{equation}
\begin{aligned}
{} [\hat{o}_{n,m},\hat{o}_{n',m'}^{\dag}]=\delta^n_{n'} \delta(m-m')
\end{aligned}
\end{equation}
Since the basis set is complete, we can decompose any arbitrary bosonic operator in the following way:
\begin{equation}
\begin{aligned}
\hat{O}= \int \mathrm{d}m \;\sum_n (O^a_{n,m}\hat{o}_{n,m}+O^b_{n,m} \hat{o}_{n,m}^{\dag}) 
\end{aligned}
\end{equation}
$O^a_{n,m}$ and $O^b_{n,m}$ are the corresponding Bogoliubov coefficients, defined in the following way:
\begin{equation}
\begin{aligned}
O^a_{n,m} & \equiv [\hat{O},\hat{o}_{n,m}^{\dag}]
\\
O^b_{n,m} & \equiv [\hat{o}_{n,m},\hat{O}] 
\end{aligned}
\end{equation}
We can conduct a transformation of basis by plugging equation (3) into (1):  
\begin{equation}
\begin{aligned}
\hat{D}_{\hat{O}}(\alpha) = \bigotimes_n \hat{D}_{\hat{o}_{n,i}}(\alpha_n)
\end{aligned}
\end{equation}
where we have defined:
\begin{equation}
\begin{aligned}
\alpha_n &\equiv \sqrt{\int \mathrm{d}m \; |O^a_{n,m}{}\alpha^* - O^b_{n,m}{}^*\alpha|^2}
\\
\alpha_n \neq 0 &\rightarrow \hat{o}_{n,i} \equiv \int \mathrm{d}m \; \left( \frac{O^a_{n,m}{}\alpha^* - O^b_{n,m}{}^*\alpha}{\alpha_n}\right) \hat{o}_{n,m}
\end{aligned}
\end{equation}
It is noted that, if $\alpha_n=0$ then $\hat{D}_{\hat{o}_{n,i}}(\alpha_n)=\mathds{1}$. Equation (5) will form the basis of the investigation of the effect of basis transformation on homodyne techniques. 
\\
\\
It is found that the contribution from $O^b_{n,m}$ distorts the amplitude and phase information of the coherent signal. As this is the contribution from the creation operator terms, we will refer to these as the negative frequency contribution throughout this paper. We find negative frequency contribution when considering non-inertial changes in reference frames. Traditional homodyne techniques do no take this effect into account, and its effect are explored in section V.
\\
\\
In this section we demonstrated a general method to transform basis for displacement operators. In the following subsection we will utilize this technique to transform a Minkowski displacement operator into a Rindler displacement operator.
\subsection{Minkowski to Rindler}
We utilize this technique to analyse how the displacement operator in the Minkowski frame transforms to the Rindler frame. A schematic map of the trajectories followed by left and right accelerated observers, Anti-Rob and Rob respectively, are shown in Fig. 1.
	\begin{figure}[h!]
		\includegraphics[width=0.45\textwidth]{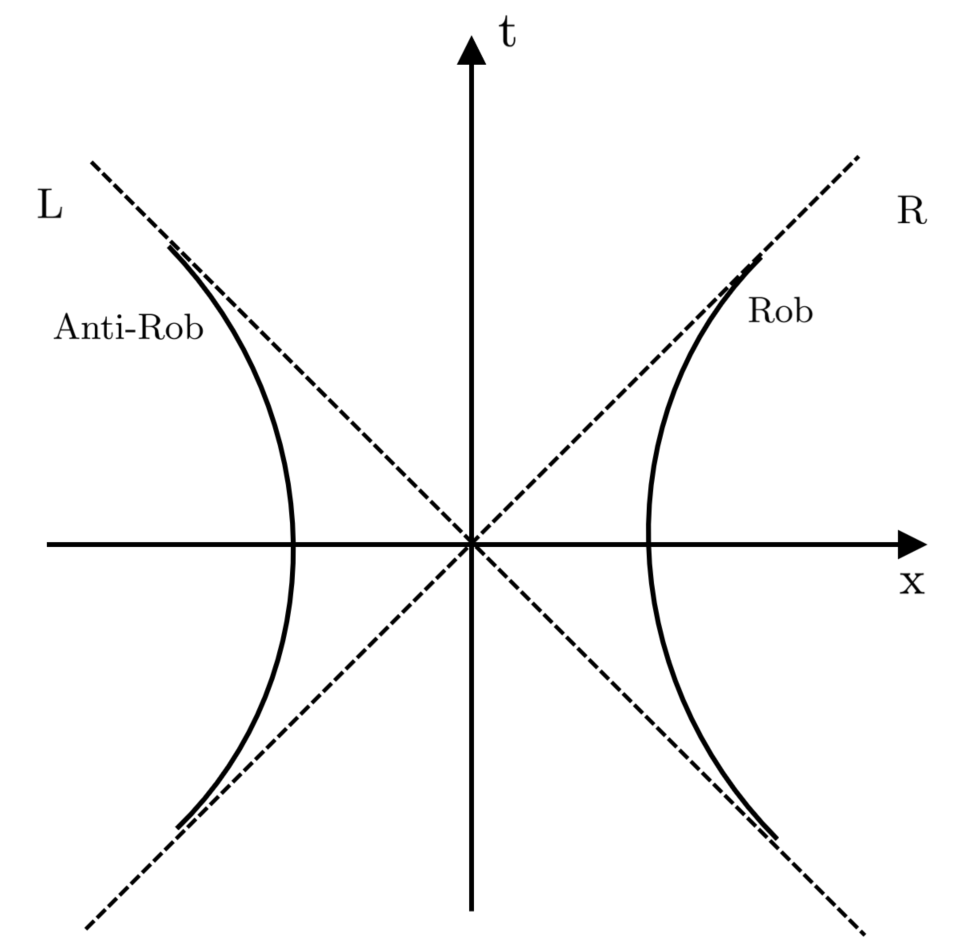}
		\caption{A (1+1) dimensional representation of the trajectories followed by an observer accelerated to the right (Rob) and left (Anti-Rob). These observers live in parts of space-time known as the right and left Rindler wedge.}	
		\label{fig: Unitary}
	\end{figure}

We introduce a normalized positive frequency Minkowski mode as follows:
\begin{equation}
\begin{aligned}
\hat{e}_f=\int_0^{\infty}\mathrm{d}k \; f(k)\hat{e}_k
\end{aligned}
\end{equation}
Where $[\hat{e}_f,\hat{e}_f^{\dag}]=1$. Through utilizing the decomposition written in equation (3), we decompose this operator into that of the Rindler frame: 
\begin{equation}
\begin{aligned}
\hat{e}_f=\int_0^{\infty}\mathrm{d}\omega \; f_{e,a}\hat{a}_{\omega}+f_{e,ac}\hat{a}_{\omega}^{\dag}+f_{e,b}\hat{b}_{\omega}+f_{e,bc}\hat{b}_{\omega}^{\dag}
\end{aligned}
\end{equation}
Through utilizing the results obtained in the Appendix A, we find that the corresponding Bogoliubov coefficients are as follows:
\begin{equation}
\begin{aligned}
f_{e,a}(\omega)& =A_{\omega} \cosh(r_{\omega})
\\
f_{e,ac}(\omega)& =-B_{\omega} \sinh(r_{\omega})
\\
f_{e,b}(\omega)& =B_{\omega} \cosh(r_{\omega})
\\
f_{e,bc}(\omega)& = - A_{\omega} \sinh(r_\omega)
\end{aligned}
\end{equation}
Where we have defined the following:
\begin{equation}
\begin{aligned}
A_{\omega}& \equiv \int_0^{\infty}\mathrm{d}k \; A_{k\omega} f(k) 
\\
B_{\omega} & \equiv \int_0^{\infty}\mathrm{d}k \; B_{k\omega} f(k)
\end{aligned}
\end{equation}
Definition of other terms can be found in the Appendix A. We now introduce the Minkowski mode displacement operator $\hat{D}_{\hat{e}_f}(\alpha_f=|\alpha_f|e^{i \phi})$. This can be transformed to the Rindler displacement operator by utilizing the transformation defined in equation (5):
\begin{equation}
\begin{aligned}
\hat{D}_{\hat{e}_f}(\alpha) = \hat{D}_{\hat{a}_{f,\phi}}(\alpha_{f,a}) \otimes \hat{D}_{\hat{b}_{f,\phi}}(\alpha_{f,b})
\end{aligned}
\end{equation}
Where we have defined the following:
\begin{equation}
\begin{aligned}
\alpha_{f,a}(\phi)& \equiv |\alpha|\sqrt{\int_0^{\infty} \mathrm{d}\omega \; |f_{e,a}(\omega)e^{-i \phi}- f_{e,ac}(\omega)^* e^{i \phi}|^2}
\\
\alpha_{f,b}(\phi)& \equiv |\alpha|\sqrt{\int_0^{\infty} \mathrm{d}\omega \; |f_{e,b}(\omega)e^{-i \phi}- f_{e,bc}(\omega)^* e^{i \phi}|^2}
\\
\hat{a}_{f,\phi} & \equiv \int_0^{\infty} \mathrm{d}\omega \left(\frac{f_{e,a}(\omega)e^{-i \phi}- f_{e,ac}(\omega)^* e^{i \phi}}{\alpha_{f,a}/|\alpha_f|} \; \hat{a}_{\omega} \right)
\\
\hat{b}_{f,\phi}& \equiv \int_0^{\infty} \mathrm{d}\omega \left(\frac{f_{e,b}(\omega)e^{-i \phi}- f_{e,bc}(\omega)^* e^{i \phi}}{\alpha_{f,b}/|\alpha_f|} \; \hat{b}_{\omega} \right)
\end{aligned}
\end{equation}
It is noticed that the phase information of the displacement operator is now carried within the operator, $\hat{a}_f$. In the limit of $|f_{e,a}(\omega)| \gg |f_{e,ac}(\omega)|$, $\hat{a}_f\propto e^{-i \phi}$. Similar methods can be taken to reverse the transformation; transform a Rindler displacement operator to a Minkowski displacement operator. %In this paper, we are interested in the effect of the horizon on homodyne detection schemes. Since the Minkowski observer does not trace any information sent by the Rindler observer, we are not interested in this scenario.
In the next section, we will look into the transformation between Rindler and delayed Rindler modes, where the delay is a constant delay with respect to Minkowski time.
\subsection{Rindler to delayed Rindler}
The delayed Rindler observers will be referred to as Anti-Rob' and Rob'. A schematic map of the trajectories followed by these observers are shown in Fig. 2.
	\begin{figure}[h!]
		\includegraphics[width=0.45\textwidth]{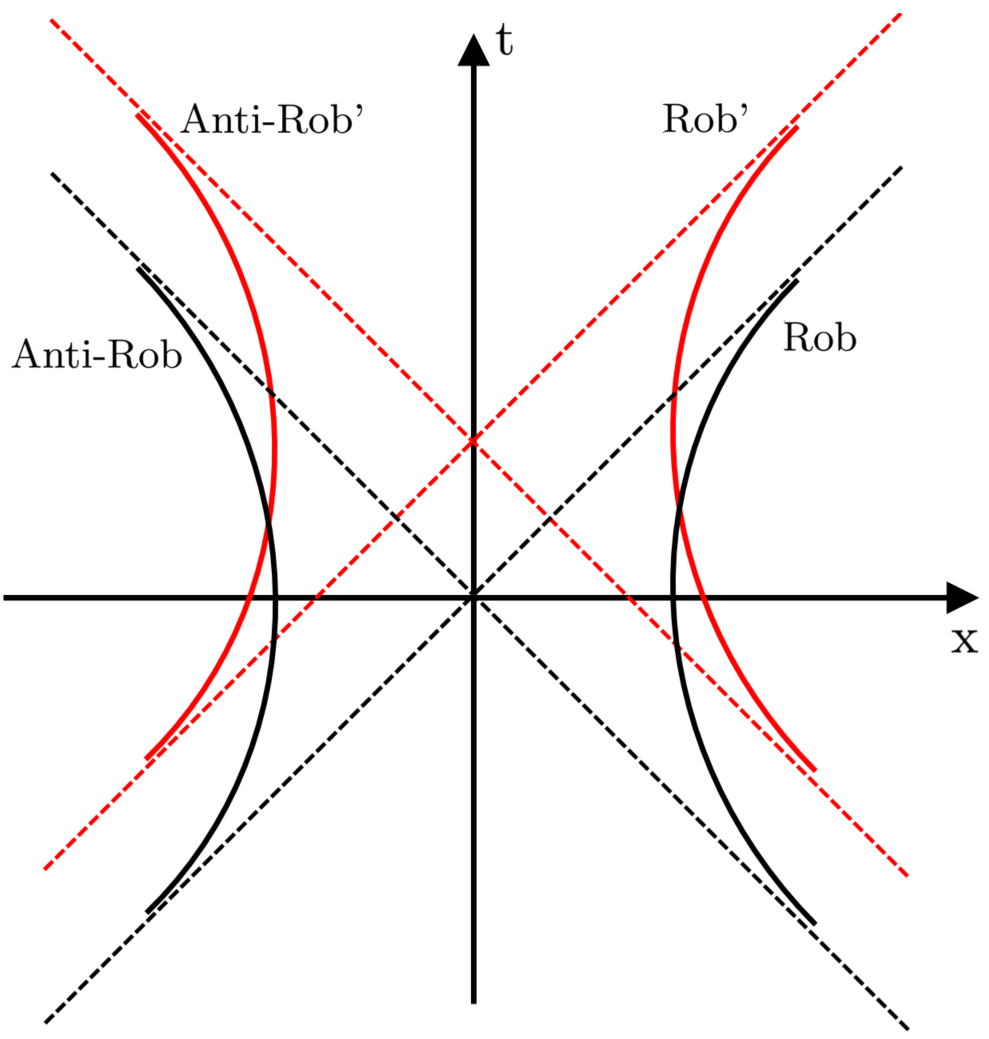}
		\caption{We write the trajectory that are followed by Rob, Anti-Rob and the Minkowski delayed Rob (Rob') and Anti-Rob  (Anti-Rob'). World lines of the red line are the ones followed by Rob' and Anti-Rob'. Anti-Rob and Rob are causally disconnected, as well as Anti-Rob' and Rob'.}
		\label{fig: Unitary}
	\end{figure}
The delayed Rindler modes are derived in Appendix B. We introduce a Rindler mode as follows:
\begin{equation}
\begin{aligned}
\hat{a}_g  = \int_0^{\infty}\mathrm{d}\omega \; g(\omega)\hat{a}_\omega
\end{aligned}
\end{equation}
We can decompose this operator to that of the delayed Rindler frame by utilizing equation (3): 
\begin{equation}
\begin{aligned}
\hat{a}_{g} = \int \mathrm{d}\omega'\; & {\alpha}^a_{\omega',g}{}^* (t) \hat{a}_{\omega'}(t)-{\beta}^a_{\omega',g}(t) \hat{a}_{\omega'}(t)^{\dag}
\\&+{\alpha}^b_{\omega',g}{}^*(t) \hat{b}_{\omega'}(t)-{\beta}^b_{\omega',g}(t) \hat{b}_{\omega'}(t)^{\dag}
\end{aligned}
\end{equation}
Where we have defined the following:
\begin{equation}
\begin{aligned}
{\alpha}^a_{\omega',g}{}^*(t) &\equiv \int_0^{\infty} \mathrm{d}\omega\; g(\omega) {\alpha}^a_{\omega',\omega}{}^*(t)
\\
{\beta}^a_{\omega',g}(t) &\equiv\int_0^{\infty} \mathrm{d}\omega\; g(\omega) {\beta}^a_{\omega',\omega}(t)
\\
{\alpha}^b_{\omega',g}{}^*(t)&\equiv \int_0^{\infty} \mathrm{d}\omega\; g(\omega){\alpha}^b_{\omega',\omega}{}^*(t)
\\
{\beta}^b_{\omega',g}(t) &\equiv \int_0^{\infty} \mathrm{d}\omega \; g(\omega) {\beta}^b_{\omega',\omega}(t)
\end{aligned}
\end{equation}
The Bogoliubov coefficients between Rindler and delayed Rindler are derived in Appendix D. We now introduce the Rindler displacement operator $\hat{D}_{\hat{a}_g}(\alpha_g=|\alpha_g|e^{i \phi})$. This can be transformed to the delayed Rindler displacement operator by utilizing the transformation defined in equation (5).
\begin{equation}
\begin{aligned}
\hat{D}_{\hat{a}_g}(\alpha_g) = \hat{D}_{\hat{a}_{g}(t)}(\alpha_{g,a(t)}) \otimes \hat{D}_{\hat{b}_{g}(t)}(\alpha_{g,b(t)})
\end{aligned}
\end{equation}
Where we have defined the following:
\begin{equation}
\begin{aligned}
\alpha_{g,a(t)}(\phi)& \equiv |\alpha_g|\sqrt{\int_0^{\infty} \mathrm{d}\omega \; |{\alpha}^a_{\omega',g}{}^* (t) e^{-i \phi}+{\beta}^a_{\omega',g}{}^*(t)  e^{i \phi}|^2}
\\
\alpha_{g,b(t)}(\phi)& \equiv |\alpha_g|\sqrt{\int_0^{\infty} \mathrm{d}\omega \; |{\alpha}^b_{\omega',g}{}^* (t) e^{-i \phi}+{\beta}^b_{\omega',g}{}^*(t)  e^{i \phi}|^2}
\\
\hat{a}_{g,\phi}(t) & \equiv \int_0^{\infty} \mathrm{d}\omega \left(\frac{{\alpha}^a_{\omega',g}{}^* (t) e^{-i \phi}+{\beta}^a_{\omega',g}{}^*(t)  e^{i \phi}}{\alpha_{g,a}/|\alpha_g|}\hat{a}_{\omega}(t) \right)
\\
\hat{b}_{g,\phi}(t)& \equiv \int_0^{\infty} \mathrm{d}\omega \left(\frac{{\alpha}^b_{\omega',g}{}^* (t) e^{-i \phi}+{\beta}^b_{\omega',g}{}^*(t)  e^{i \phi}}{\alpha_{g,b}/|\alpha_g|} \hat{b}_{\omega}(t)\right)
\end{aligned}
\end{equation}
In this section we introduced the universal transformation of displacement operators. We then demonstrated how to use this technique by utilizing it in two specific cases. 
\\
\\
We will gain insight into how homodyne detection schemes work in differing reference frames by utilizing the universal transformation of displacement operators. In the following section we introduce self-homodyne and balanced-homodyne detection schemes.
\section{Traditional Homodyne Techniques}
We consider a scenario where a signaller creates a Gaussian signal by applying a Gaussian operation, $\hat{U}$, onto the initial state. The observer is interested in extracting the QI of this Gaussian signal. Homodyne tomography \cite{Lvovsky2009} can be utilized to characterise the QI (Wigner function) of a particular field mode. We introduce an arbitrary bosonic basis set, $\hat{O}_i$, that is complete for the signaller (it does not necessarily need to be a set that is complete globally); $[\hat{O}_i,\hat{O}_{i'}^{\dag}]=\delta^i_{i'}$. We denote the annihilation operator of the field mode that is of interest as $\hat{O}_f$.
\\
\\
 Since the created signal is a Gaussian state, the analysis of the first and second order moment \cite{Weedbrook2012} is sufficient to characterise the Wigner function of a particular output mode \cite{Scully1997}. For non-Gaussian interactions, higher order quadrature moments must be analysed to obtain the full QI of the mode. In the following section, we will explain how balanced-homodyne detection scheme can be implemented to extract the QI of the mode $\hat{O}_f$.
\subsection{Balanced-Homodyne}
Balanced homodyne detection is a well-known technique in quantum optics. Here we generalize this technique to situation where the signaller and observer are in different reference frames. In balanced-homodyne detection scheme, the signaller couples $\hat{O}_f$ with a strong coherent local oscillator. To do this, we introduce another complete (for the signaller) bosonic basis set for the local oscillator $\hat{O}_{L,i}$; $[\hat{O}_{L,i},\hat{O}_{i'}]=[\hat{O}_{L,i},\hat{O}_{i'}^{\dag}]=0 \; \forall i,i'$. The strong coherent signal can be created in the basis set of the local oscillator by applying the displacement operator, $\hat{D}_{L,f}(|\alpha|)\equiv \exp[|\alpha|(\hat{O}_{f,L}^{\dag}-\hat{O}_{f,L})]$.
\\
\\
We introduce an arbitrary basis set $\hat{o}_i$ which completes the basis set for an observer that is receiving the signal. We can analyse what is observed by this observer by evolving this basis set via the Heisenberg picture. The state with the Gaussian signal and large coherent local oscillator can be created by acting the Gaussian unitary operator, $\hat{U}_s$, onto the initial state. In the Heisenberg picture, we interpret this as the following:
	\begin{equation}
	\begin{aligned}
	\hat{o}_{i}' & \equiv \hat{U}_S^{\dag}\hat{o}_{i} \hat{U}_S
	\\
	\hat{U}_S &\equiv \hat{U}\otimes \hat{D}_{L,f}(|\alpha|)
	\end{aligned}
	\end{equation}
In balanced homodyne detection, the observer applies a tunable phase shift ($\hat{U}_{\phi}$) onto the local oscillator, followed by a balanced beam splitter ($\hat{U}_{BS}$) which acts on all incoming modes. 
	\begin{equation}
	\begin{aligned}
	\hat{U}_{\phi}&\equiv \exp[i \phi \sum_i \hat{o}_{L,i}^{\dag}\hat{o}_{L,i}]
	\\
	\hat{U}_{BS}& \equiv \exp[\frac{\pi}{2} \sum_i (\hat{o}_{i}^{\dag}\hat{o}_{L,i}-\hat{o}_{L,i}^{\dag}\hat{o}_{i})]
	\\
	\hat{U}_O\ & \equiv \hat{U}_{BS} \hat{U}_\phi	
	\end{aligned}
	\end{equation}
In the Heisernberg picture, the operator evolves in the following way:
\begin{equation}
\begin{aligned}
	\hat{o}_{i}'' & \equiv \hat{U}_{S}^{\dag}\hat{U}_{O}^{\dag} \hat{o}_{i}\hat{U}_{O}\hat{U}_{S}	
\end{aligned}
\end{equation}
The quadrature amplitude and variance of the mode $\hat{O}_f$ can then be computed by utilizing the following definitions:
\begin{equation}
\begin{aligned}
X_{f,b}(\phi) & \equiv \frac{\braket{\hat{N}_L'' - \hat{N}''}}{\sqrt{\braket{\hat{N}_L'}}}
\\
V_{f,b}(\phi) & \equiv \frac{\braket{(\hat{N}_L'' - \hat{N}'')^2}-\braket{\hat{N}_L'' - \hat{N}''}^2}{\braket{\hat{N}_L'}}
\end{aligned}
\end{equation}
Where we have introduced:
\begin{equation}
\begin{aligned}
\hat{N} & \equiv \sum_i \hat{o}^{\dag}_i{} \hat{o}_i{}
 \\
\hat{N}_L & \equiv \sum_i \hat{o}^{\dag}_{L,i}{} \hat{o}_{L,i}{}
\end{aligned}
\end{equation}
and their superscript versions. A schematic map of the process involved for balanced homodyne detection is demonstrated in Fig. 3. 	\begin{figure}[h!]
\includegraphics[width=0.45\textwidth]{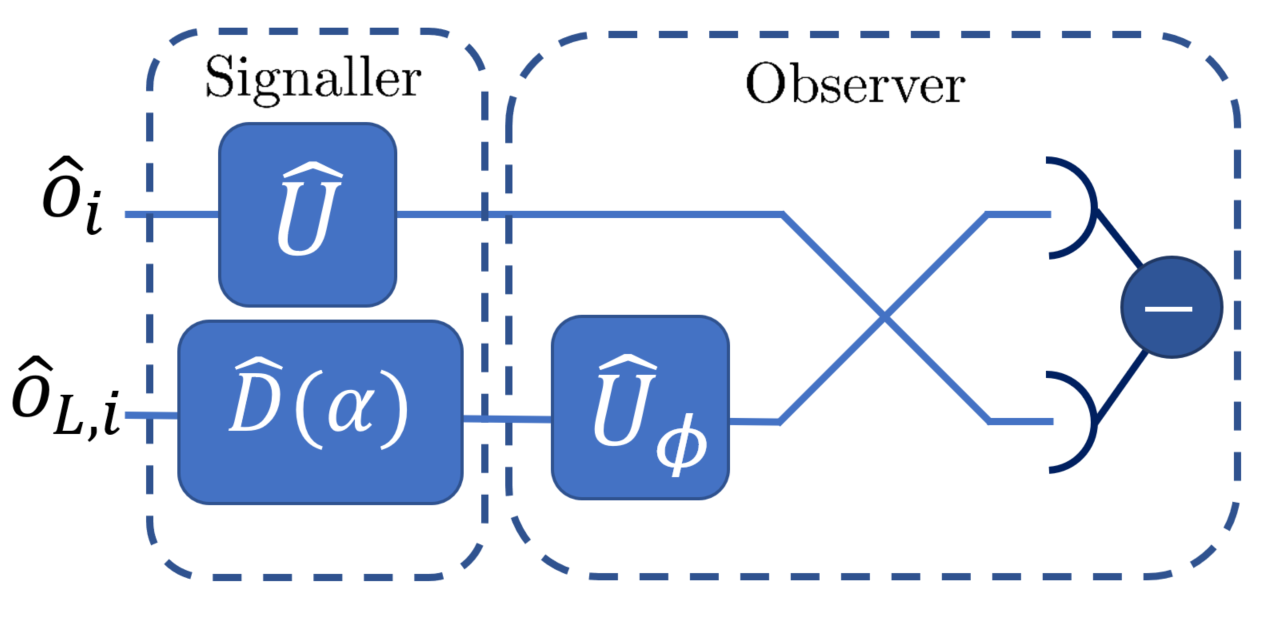}
		\caption{A schematic diagram which demonstrates how balanced homodyne detection works.}	
		\label{fig: Unitary}
	\end{figure}
\\
\\
It is easy to show the following:
	\begin{equation}
	\begin{aligned}
\hat{N}_L''-\hat{N}''	& =\sum_i(\hat{o}_{L,i}^{\dag}{}' \hat{o}_{i}{}' \; e^{-i \phi} + \hat{o}_{i}^{\dag}{}' \hat{o}_{L,i} {}'\; e^{i \phi})
	\end{aligned}
	\end{equation}
We now restrict to the case where the signaller and observer are both initially in the vacuum state and in the same reference frame. In this regime we explain why equation (21) is valid. In this regime, the set $\hat{o}_i$ and $\hat{O}_i$ coincides with each other, thus $\braket{\hat{o}_{L,i}'}=\delta^i_f |\alpha|$, as a result we find the following:
\begin{equation}
\begin{aligned}
X_{f,b}(\phi) & = \braket{\hat{o}_{f}{}' \; e^{-i \phi} + \hat{o}_{f}^{\dag}{}'  e^{i \phi}}
\end{aligned}
\end{equation}
hence $X_f=\braket{\hat{X}_f}$, as required. Furthermore, by utilizing the property that $\braket{\hat{o}_{L,i}'\hat{o}_{L,i'}'}=\delta^i_f\delta^{i'}_f |\alpha|^2$ and $\braket{\hat{o}_{L,i}'\hat{o}_{L,i'}'^{\dag}}=\delta^i_f\delta^{i'}_f |\alpha|^2+\delta^i_{i'}$
\begin{equation}
	\begin{aligned}
\braket{(\hat{N}_L''-\hat{N}'')^2 }& = |\alpha|^2 \braket{(\hat{o}_{f}{}' \; e^{-i \phi} + \hat{o}_{f}^{\dag}{}'  e^{i \phi})^2}+\braket{\hat{N}'}
	\end{aligned}
	\end{equation}
Utilizing this equation, we find that $V_{f,b}(\phi)$ simplifies to the following:
\begin{equation}
\begin{aligned}
V_{f,b}(\phi) & =\braket{\hat{V}_f}+{\braket{\hat{N}'}}/|\alpha|^2
\end{aligned}
\end{equation}
The variance that is found via the balanced homodyne detection scheme is valid when we set the coherent signal of the local oscillator much larger than the number of particles created via the unitary. 
%In this paper, we are interested in whether balanced homodyne detection is still valid when the reference frame of the sender and the observer do not coincide with each other.
\subsection{Self-Homodyne}
In this section we introduce the self-homodyne detection scheme. 
%Balanced-homodyne is conducted through displacing the local oscillator which is coupled to the mode of interest. This allowed the detector to put and arbitrary phase onto local oscillator without affecting the Gaussian signal. 
In self-homodyne, we directly displace the mode that is of interest. The phase information is encoded within the displacement operator (see Fig. 4).
%. As a result, self homodyne detection scheme requires the phase reference to be encoded by the signaller. A schematic representation of the set up of self-homodyne detection is demonstrated in Fig. 4.
	\begin{figure}[h!]
\includegraphics[width=0.45\textwidth]{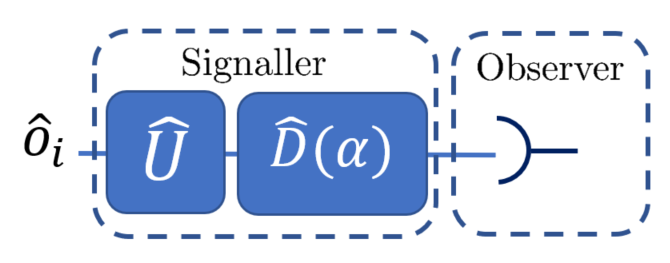}
		\caption{A schematic diagram which demonstrates how self homodyne detection works.}	
		\label{fig: Unitary}
	\end{figure}
\\
\\
We begin by creating the Gaussian signal as follows:
	\begin{equation}
	\begin{aligned}
	\hat{o}_{i}' & \equiv \hat{U}^{\dag} \hat{o}_{i}\hat{U}
	\end{aligned}
	\end{equation}
We then	introduce the annihilation operator of the mode that is of interest as $\hat{O}_f$. Then the displacement operator is defined as $\hat{D}_f(\alpha=|\alpha| e^{i \phi})=\exp[\alpha \hat{O}_f^{\dag}-\alpha^* \hat{O}_f]$. The signaller then couples the signal with a strong coherent signal. In the Heisernberg picture, this is interpreted in the following way:
	\begin{equation}
	\begin{aligned}
	\hat{o}_{i}'' & \equiv \hat{U}^{\dag} \hat{D}^{\dag}_f(\alpha)\hat{o}_{i}\hat{D}_f(\alpha)\hat{U}
	\end{aligned}
	\end{equation}
The quadrature amplitude and variance of $\hat{O}_f$ can be computed by comparing the particle count of an output with and without the signal for various $\phi$. They are computed through utilizing the following equations \cite{Onoe2018}:
\begin{equation}
\begin{aligned}
X_{f,s}(\phi) &=\left(\braket{\hat{N}''}-\braket{\hat{N_0}}\right)/{\sqrt{\braket{\hat{N}_0}}} 
\\
V_{f,s}(\phi) & = \left(\braket{(\hat{N}''{})^2}-\braket{\hat{N}''}^2\right)/\braket{\hat{N}_0}
\end{aligned}
\end{equation}
Where we have defined the following:
\begin{equation}
\begin{aligned}
\hat{N} & \equiv \sum_i \hat{o}^{\dag}_i{} \hat{o}_i{}
 \\
\hat{N}_0 &\equiv \hat{D}_f^{\dag}(\alpha)\left(\sum_i \hat{o}^{\dag}_i{}' \hat{o}_i{}\right)\hat{D}_f^{\dag}(\alpha)
\end{aligned}
\end{equation}
A more rigorous derivation, with explanation of the regimes when these equations are valid are conducted in \cite{Onoe2018}.
\\
\\
In the next section we develop a complete homodyne detection scheme which is valid for communication between differing reference frames. 
%We then explore the differences between the new and traditional methods to highlight the issues with the traditional methods.
\section{Ideal Homodyne Tomography}
The issue with traditional homodyne techniques is that the measurement and signal basis are different. This can lead to apparent decoherence effects \cite{Daiqin2017b, Onoe2018}. By utilizing the circuit model \cite{Daiqin2017a}, we can reproduce a system where all interactions are effectively occurring in the observer's reference frame. The operational set-up is shown in Fig. 5.
	\begin{figure}[h!]
\includegraphics[width=0.45\textwidth]{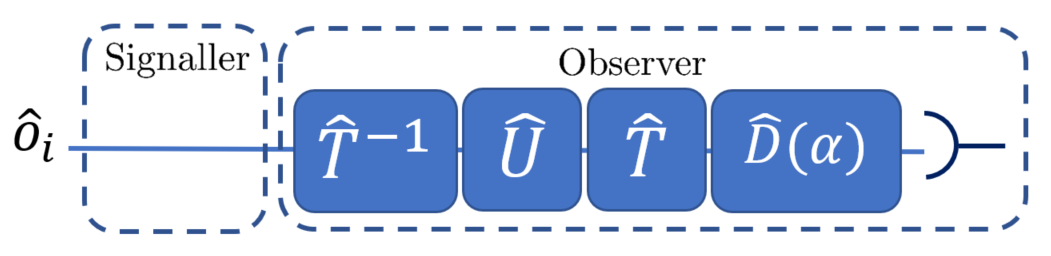}
		\caption{The unitary has been moved to the frame of the observer by introducing the basis transformation operator. In this decomposition, all interaction are conducted in one reference frame, thus the the standard self-homodyne detection scheme can be implemented.}	
		\label{fig: Unitary}
	\end{figure}
	\\
\\
The circuit model \cite{Daiqin2017a} uses the fact that a unitary interaction can be transformed to a unitary interaction in a different frame by introducing a basis transformation operator. This unitary basis transformation operator is denoted as $\hat{T}$. Fig 5. has utilized the circuit model to move the interaction to the observer's frame. The cost of this method is that the observer understands the modal decomposition of the interaction and can produce a coherent signal $\hat{D}(\alpha)$ which may have a very complex modal structure.
\\
\noindent
Alternatively we can impose an operational constraint that the observer does not know the modal decomposition of the interaction. This means that the displacement operator must be created by the signaller. We utilize the universal transformation for displacement operators to move the displacement operator into the signaller's reference frame. Fig. 6 is the ideal homodyne technique with this operational constraints.
\\
\\ 
We note that, during the transformation of reference frame, the displacement operator may have increased to more than one displacement operator. Nevertheless, the essence of ideal-homodyne tomography is captured within Fig. 6. We note that the mathematical set-up is identical to that of the self-homodyne detection scheme. 
	\begin{figure}[h!]
\includegraphics[width=0.45\textwidth]{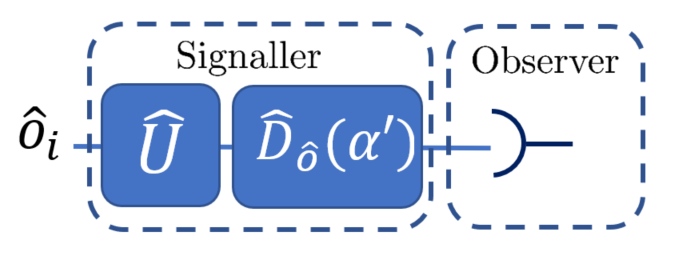}
		\caption{Utilizing the technique of the circuit model and universal transformation of displacement operators, we can move all the operators to the signaller's reference frame.}	
		\label{fig: Unitary}
	\end{figure}
\\
\noindent
The cost of utilizing this detection scheme is that the signaller must know in which frame the observer will be. Utilizing this information they must deduce the required modal decomposition of the reference signal $\hat{D}(\alpha)'$, which may have a very complex modal structure. Furthermore, in some scenarios the displacement operator $\hat{D}(\alpha)'$ may require the signaller to have access to a part of space-time which is space-like separated from them. Due to these reasons ideal homodyne may be impractical in certain situations, however it clarifies the information which is lost when traditional homodyne detection methods are utilized. Hence it will still be a useful theoretical tool in understanding the full quantum information of signals and help explain obscure effects that are observed utilizing traditional homodyne schemes.
\section{Coherent Signalling between different reference frames}
In the previous section we developed the ideal homodyne technique. This homodyne technique was developed because traditional homodyne technique cannot always communicate the full QI of the signal to a observer in an different reference frame. 
\\
\\
In this section we will compare the three homodyne detection schemes by considering the simplest form of quantum communication; a coherent signal. We will consider two cases; an inertial Minkowski observer sending a coherent signal to an accelerated observer and an accelerated observer sending a coherent signal to a delayed Rindler observer. In section A, we obtain analytical expressions for quadrature amplitude and variance for the two cases. We then conclude by commenting on the differences between the three detection schemes. In section B we give numerical results by considering a Gaussian wave-packet mode signal, in interesting regimes.
\subsection{General Case}
\subsubsection{Minkowski to Rindler}
In this section we consider the case when a Minkowski observer sends a coherent signal to Anti-Rob. Su \textit{et al.} \cite{Daiqin2014} analysed a similar setting utilizing balanced-homodyne detection. We set the signal that is sent by the inertial observer to be the following:
\begin{equation}
\hat{U}\equiv \hat{D}_f(\beta=|\beta|e^{i \psi})
\end{equation}
We utilize the technique discussed in section II to transform basis. This signal then transforms to the following:
\begin{equation}
\hat{U}= \hat{D}_{\hat{a}_{f,\psi}}(\beta_{f,a}) \otimes \hat{D}_{\hat{b}_{f,\psi}}(\beta_{f,b})
\end{equation}
We begin by analysing the situation via the balanced homodyne detection method. The unitary involved are the following in the Rindler frame:
	\begin{equation}
	\begin{aligned}
	\hat{U}_{\phi}&\equiv \exp[i \phi \int \mathrm{d}\omega\; \hat{b}_{L,\omega}^{\dag}\hat{b}_{L,\omega}]
	\\
	\hat{U}_{BS}& \equiv \exp[\frac{\pi}{2} \int \mathrm{d}\omega (\hat{b}_{\omega}^{\dag}\hat{b}_{L,\omega}-\hat{b}_{L,\omega}^{\dag}\hat{b}_{\omega})]
	\\
	\hat{D}_{L,f}(|\alpha|)&= \hat{D}_{\hat{a}_{f,0}}(\alpha_{f,a}) \otimes \hat{D}_{\hat{b}_{f,0}}(\alpha_{f,b})
\end{aligned}
\end{equation}
The number operator in the right Rindler frame is defined as follows:
\begin{equation}
\begin{aligned}
\hat{N} & = \int \mathrm{d}\omega \; \hat{b}_{\omega}^{\dag}\hat{b}_{\omega}
\\
\hat{N}_L & = \int \mathrm{d}\omega \; \hat{b}_{L,\omega}^{\dag}\hat{b}_{L,\omega}
\end{aligned}
\end{equation}
Following the process discussed in the previous section, we find that the quadrature amplitude can be calculated utilizing the following: 
\begin{equation}
\begin{aligned}
X_{f,b}(\phi) & = \frac{\int \mathrm{d}\omega\; \braket{\hat{b}_{L,\omega}^{\dag}{}' \hat{b}_{\omega}{}' \; e^{-i \phi} + \hat{b}_{\omega}^{\dag}{}' \hat{b}_{L,\omega} {}'\; e^{i \phi}}
}{\sqrt{\braket{\hat{N}_L'}}}
\end{aligned}
\end{equation}
Utilizing the circuit model analysis \cite{Daiqin2017a}, we find that the operators evolve under the displacement operator in the following way:
\begin{equation}
\begin{aligned}
\hat{b}_{\omega}' & =\hat{b}_{\omega}+\beta_{f,\hat{b}_{\omega}}(\psi)
\\
\hat{b}_{L,\omega}' & =\hat{a}_{L,\omega}+\alpha_{f,\hat{b}_{\omega}}(0)
\\
\alpha_{f,\hat{a}_{\omega}}(\phi) & \equiv |\alpha_f| \left(f_{e,b}(\omega)^*e^{i \phi}- f_{e,bc}(\omega)  e^{-i \phi} \right)
\\
\beta_{f,\hat{b}_{\omega}} (\psi)& \equiv |\beta_f| \left(f_{e,b}(\omega)^*e^{i \psi}- f_{e,bc}(\omega) e^{-i \psi}  \right)
\end{aligned}
\end{equation}
 As a result, the quadrature amplitude and variance that is found via the balanced homodyne detection method is as follows:
\begin{equation}
\begin{aligned}
X_{f,b}(\phi) & = \frac{\int \mathrm{d}\omega\; 2Re[\alpha_{f,\hat{b}_{\omega}(0)}^* \beta_{f,\hat{b}_{\omega}}(\psi) \; e^{-i \phi}]
}{\alpha_{f,b}(0)}
\\
V_{f,b}(\phi) & \approx \int_0^{\infty} \mathrm{d}\omega \left(\left|\frac{\alpha_{f,\hat{b}_{\omega}}(0)}{\alpha_{f,b}(0)}\right|^2 \; (1+2 \sinh(r_\omega)^2) \right)
\end{aligned}
\end{equation}
Self-homodyne detection scheme can be conducted by coupling the signal with a strong coherent signal:
\begin{equation}
	\begin{aligned}
	\hat{D}_{f}(\alpha=|\alpha|e^{i \phi})&= \hat{D}_{\hat{a}_{f,\phi}}(\alpha_{f,a}) \otimes \hat{D}_{\hat{b}_{f,\phi}}(\alpha_{f,b})
\end{aligned}
\end{equation}
We then find that the quadrature amplitude and variance found via the self homodyne detection method are:
\begin{equation}
\begin{aligned}
X_{f,s}(\phi) & = \frac{\int \mathrm{d}\omega\; 2Re[\alpha_{f,\hat{b}_{\omega}}(\phi)^* \beta_{f,\hat{b}_{\omega}}(\psi) ]}{\alpha_{f,b}(\phi)}
\\
V_{f,s}(\phi) & \approx \int_0^{\infty} \mathrm{d}\omega \left(\left|\frac{\alpha_{f,\hat{b}_{\omega}}(\phi)}{\alpha_{f,b}(\phi)}\right|^2 \; (1+2 \sinh(r_\omega)^2) \right)
\end{aligned}
\end{equation}
The ideal homodyne detection scheme can be conducted via utilizing the displacement operator $\hat{D}_{\hat{a}_{f,\phi}}(\beta_{f,a}e^{i\psi})$. Following similar steps, we find the following: 
\begin{equation}
\begin{aligned}
X_{f,I}(\phi) & = \frac{\int \mathrm{d}\omega\; 2 Re[\alpha_{f,\hat{b}_{\omega}}(\psi)^* \beta_{f,\hat{b}_{\omega}}(\psi)e^{-i \phi}]}{\alpha_{f,b}(\psi)}
\\
V_{f,I}(\phi) & \approx \int_0^{\infty} \mathrm{d}\omega \left(\left|\frac{\alpha_{f,\hat{b}_{\omega}}(\psi)}{\alpha_{f,b}(\psi)}\right|^2 \; (1+2 \sinh(r_\omega)^2) \right)
\end{aligned}
\end{equation}
In the next subsection we find the quadrature amplitude and variance of a coherent signal that is sent from a Rindler observer to a delayed Rindler observer. We will analyse the difference between these homodyne technique later in this section.
\subsubsection{Rindler to Minkowski delayed Rindler}
In this subsection we consider the case where Rob sends a coherent signal to Anti-Rob'. Su \textit{et al.} \cite{DaiqinThesis} analysed a similar setting utilizing balanced-homodyne detection. The corresponding number operator is as follows:
\begin{equation}
\begin{aligned}
\hat{N}&= \int \mathrm{d}\omega \; \hat{b}_{\omega}^{\dag}(t)\hat{b}_{\omega}(t)
\end{aligned}
\end{equation}
We set the signal that is sent by the Rindler observer to be the following:
\begin{equation}
\begin{aligned}
\hat{U} & \equiv \hat{D}_g(\beta=|\beta|e^{i \psi})
\end{aligned}
\end{equation}
%\\ & = \hat{D}_{\hat{a}_{g,\psi}(t)}(\beta_{g,a(t)}(\psi)) \otimes \hat{D}_{\hat{b}_{g,\psi}(t)}(\beta_{g,b(t)}(\psi))
%We begin by introducing the unitary involved for balanced homodyne detection method:
%	\begin{equation}
%	\begin{aligned}
%	\hat{U}_{\phi}&\equiv \exp[i \phi \int \mathrm{d}\omega\; \hat{b}_{L,\omega}^{\dag}(t)\hat{b}_{L,\omega}(t)]
%	\\
%	\hat{U}_{BBS}& \equiv \exp[\frac{\pi}{2} \int \mathrm{d}\omega (\hat{b}_{\omega}^{\dag}(t)\hat{b}_{L,\omega}(t)-\hat{b}_{L,\omega}^{\dag}(t)\hat{b}_{\omega}(t))]
%	\\
%	\hat{D}_{L,g}(|\alpha|)&= \hat{D}_{\hat{a}_{g,0}(t)}(\alpha_{g,a}(t)) \otimes \hat{D}_{\hat{b}_{g,0}(t)}(\alpha_{g,b}(t))
%\end{aligned}
%\end{equation}
%The unitary involved for self homodyne detection scheme is as follows:
%\begin{equation}
%	\begin{aligned}
%	\hat{D}_{g}(\alpha=|\alpha|e^{i \phi})&= \hat{D}_{\hat{a}_{g,\phi}(t)}(\alpha_{g,a(t)}) \otimes \hat{D}_{\hat{b}_{g,\phi}(t)}(\alpha_{g,b(t)})
%\end{aligned}
%\end{equation}
%For both detection schemes, the number operator in the left Minkowski delayed Rindler frame is defined as follows:
Following similar process to the previous subsection, we find that the quadrature amplitude and variance that are found via balanced, self and ideal homodyne are as follows:
\begin{equation}
\begin{aligned}
X_{g,b}(\phi) & = \frac{\int \mathrm{d}\omega\; 2Re[\alpha_{g,\hat{b}_{\omega}(t)}^*(0) \beta_{g,\hat{b}_{\omega}(t)}(\psi)  e^{-i \phi}]
}{\alpha_{g,b(t)}(\psi)}
\\
V_{g,b}(\phi) & \approx \int_0^{\infty} \mathrm{d}\omega \left(\left|\frac{\alpha_{g,\hat{b}_{\omega}(t)}(0)}{\alpha_{g,b(t)}(0)}\right|^2 \; (1+2 \sinh(r_\omega)^2) \right)
\end{aligned}
\end{equation}
\begin{equation}
\begin{aligned}
X_{g,s}(\phi) & = \frac{\int \mathrm{d}\omega\; 2 Re[\alpha_{g,\hat{b}_{\omega}(t)}(\phi)^* \beta_{g,\hat{b}_{\omega}(t)}(\psi)] }{\alpha_{g,b(t)}(\phi}
\\
V_{g,s}(\phi) & \approx \int_0^{\infty} \mathrm{d}\omega \left(\left|\frac{\alpha_{g,\hat{b}_{\omega}(t)}(\phi)}{\alpha_{g,b(t)}(\phi)}\right|^2 \; (1+2 \sinh(r_\omega)^2) \right)
\end{aligned}
\end{equation}
\begin{equation}
\begin{aligned}
X_{g,I}(\phi) & = \frac{\int \mathrm{d}\omega\; \alpha_{g,\hat{b}_{\omega}(t)}(\psi)^* \beta_{g,\hat{b}_{\omega}(t)}(\psi)e^{-i\phi}}{\alpha_{g,b(t)}}
\\
V_{g,I}(\phi) & \approx \int_0^{\infty} \mathrm{d}\omega \left(\left|\frac{\alpha_{g,\hat{b}_{\omega}(t)}(\psi)}{\alpha_{g,b(t)}(\psi)}\right|^2 \; (1+2 \sinh(r_\omega)^2) \right)
\end{aligned}
\end{equation}
Where we have defined:
\begin{equation}
\begin{aligned}
\alpha_{g,\hat{b}_{\omega}(t)}(\phi) & \equiv |\alpha_g| \left(\alpha^b_{\omega,g}(t)e^{i \phi}+ \beta^b_{\omega,g}(t) e^{-i \phi}  \right)
\\
\beta_{g,\hat{b}_{\omega}(t)} (\psi)& \equiv |\beta_g| \left(\alpha^b_{\omega,g}(t)e^{i \psi}+ \beta^b_{\omega,g}(t) e^{-i \psi}  \right)
\end{aligned}
\end{equation}
%We find that the expression found in equation (53), (55) and (60) are quite similar. We make two key observations. One is how the phase information is distorted during the transformation of reference frame. This distortion can be understood as an artifact of positive frequency mode in one reference frame transforming into both positive and negative frequency mode in another frame. In the case of Minkowski to Rindler, a positive frequency Rindler mode (e.g. $\hat{a}_{\omega}$) is decomposed into both positive frequency (i.e. $\hat{e}_{k}$, etc.) and negative frequency (i.e. $\hat{e}_{k}^{\dag}$, etc.) Minkowski modes. 
%\\
%\\
%The phase information is distorted between the two reference frames when the contribution of the negative frequency mode is large. For self homodyne, the phase reference is sent from the signaller, and hence the phase information will not be processed like that of a ordinary homodyne method. 
%\\
%\\
\subsubsection{Comparing Homodyne Techniques}
There are three main effects of transformation of basis for displacement operators. The most obvious effect is that there is not a one to one correspondence of the phase. This is to say that in general, $\alpha(\phi)' \neq \alpha'(0) e^{i \phi}$. The second is that the phase affects the modal shape in the new reference frame. Lastly, the phase can also affect the amplitude of the signal in the new reference frame. We note that $\psi$ is the phase of the signal and is a constant, while $\phi$ is the phase of the detection scheme and is a free parameter.
We will analyse the effect of $\phi$ on the quadrature amplitude and variance in this section.
\\
\\
Self-homodyne encodes the phase information in the signaller's reference frame, as a result we have a $\alpha(\phi)'$ term for the quadrature amplitude. This is an issue, as we cannot always write $\alpha(\phi)' \approx \alpha'e^{\i \phi}$. As a result when self homodyne detection is implemented, the quadrature amplitude may not be sinusoidal with $\phi$. For balanced homodyne detection, this means that the phase difference between the signal, $\beta(\psi)'$, and the reference signal $\alpha(0)'$ is not always $\psi$. As a result the relative phase difference between the signal and the coherent signal is not preserved with the change in reference frame.
\\
\\The second effect means that the a phase difference between the signal and reference signal leads to a smaller overlap between the signal and the reference mode. For balanced homodyne, the overlap between $\alpha(0)'$ and $\beta(\psi)'$ is maximised when $\psi=0$. All other cases leads to an observed quadrature amplitude which is smaller than the actual value. For self-homodyne, the amount of overlap changes with $\phi$. The two modes completely overlap with each other when $\phi=\psi + n \pi, \forall n \in \mathbb{Z} $. This corresponds to when self-homodyne gives the maximum quadrature amplitude readings, and is the special case when the ideal and self-homodyne detection scheme coincides with each other.
\\
\\
The last effect simply implies that the normalisation constant for self-homodyne is not constant with $\phi$, which is naturally accounted for in the formalism of the homodyne techniques. Thus, there are no significant issues due to this effect. 
\\
\\
We now analyse how these effects affect the observed variance. The variance is only affected by the phase's effect on modal structure. The accelerated observers see a thermal background. The reference signal tells the observer which thermal background to analyse. The phase difference between the signal and reference signal leads to the observer analysing a different thermal background. Balanced homodyne will lead to the observer analysing the thermal background corresponding to the wavepacket mode with the phase $0$. Self homodyne will lead to the observer analysing the thermal background corresponding to the wavepacket mode with phase $\phi$, hence the variance readings will change with $\phi$. Traditional methods are valid when the two reference frames coincide as the phase difference between the two modes does not influence the modal structure that is being analysed.
%In traditional homodyne schemes, the phase difference between the signal and the reference signal does not affect which mode is being analysed by the observer. 
\\
\\
We will see these effects explicitly in the next sections, as we go through numerical examples.
\subsection{Gaussian Wave-packet Mode}
\subsubsection{Minkowski to Rindler}
To conduct numerical analysis, we must define the wavepacket mode of $\hat{e}_f$. In this paper we will consider a generalized Gaussian wavepacket mode. The wave-packet mode is defined as follows:
	\begin{equation}
	\begin{aligned}
	f(k;k_0,\sigma,V_0)& =A\sqrt{k}(\frac{1}{2\pi \sigma^2})^{1/4} \exp[-\frac{(k-k_0)^2}{4\sigma^2}-ikV_0] 
	\end{aligned}
	\end{equation}
	Where $A$ is the normalization constant such that $\int |f(k)|^2 \mathrm{d}k = 1$. When $0.4 k_0 > \sigma$, $A \approx 1/\sqrt{k_0}$ and this wave-packet mode has a field that is of the standard Gaussian form. Analysis shows that the phase-information removed at the horizon distorts the QI sent from Minkowski to Rindler frame. The effect was optimized when the local phase of the wave-packet was $\pi/2 +n \pi, \; \forall n \in \mathbb{Z}$ at the horizon and minimized when the local phase was $n \pi, \; \forall n \in \mathbb{Z}$. These effects diverge if we do not set a low frequency cut-off. Hence, we have introduced a low frequency cut-off, $\omega_{min}$, to the detector for these plots. In Fig. 7, we see the direct effect of this on the variance.
\begin{figure}[h!]
\includegraphics[width=0.45\textwidth]{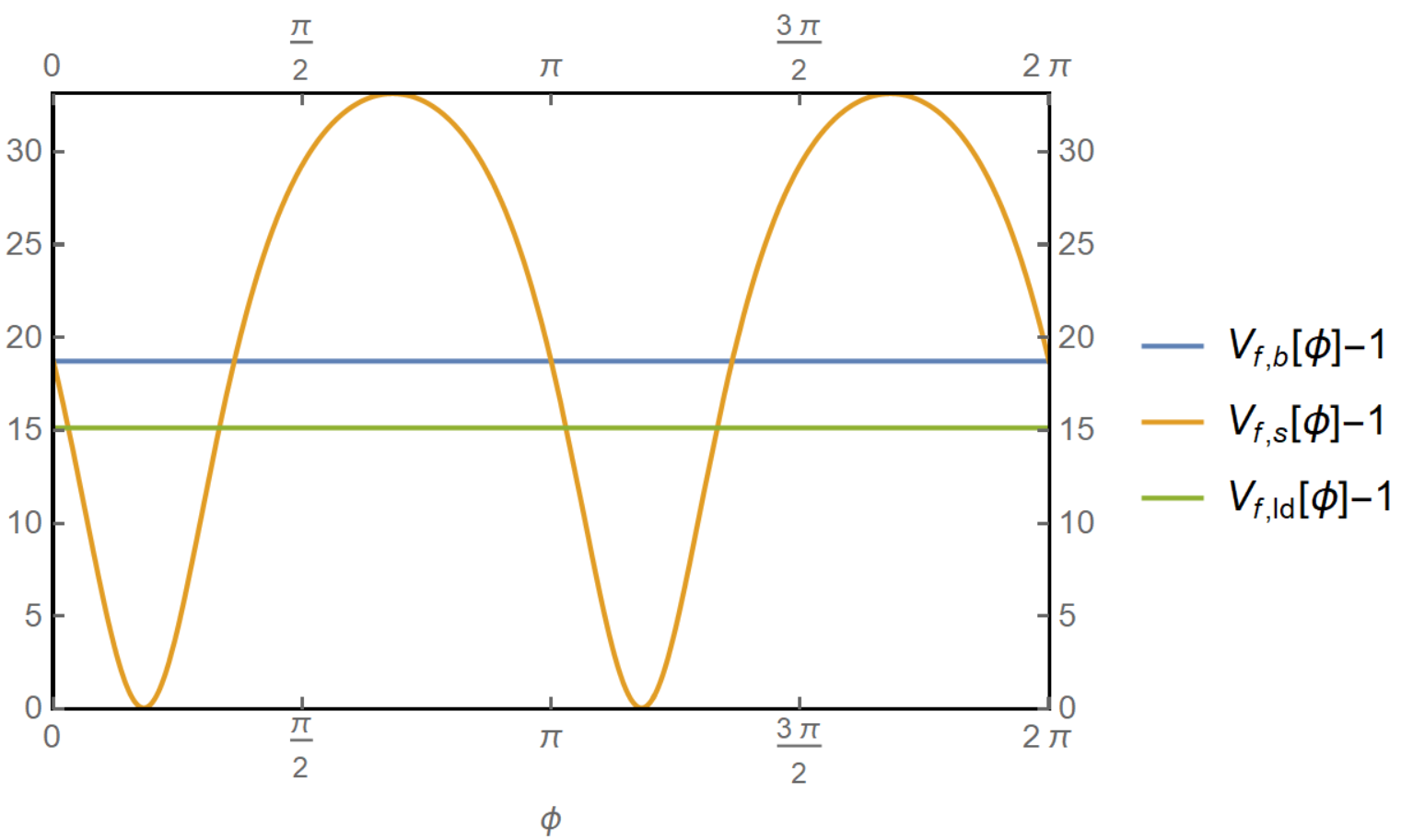}
		\caption{A plot which demonstrates how the local phase at the horizon affects the variance. We have utilized the following settings: $a=1,\; V_0=1,\; k_0=1, \; \sigma=0.2, \; |\beta_f|=1, \psi= \pi/3, \; \omega_{min}=10^{-3}$.}	
		\label{fig: Unitary}
	\end{figure}
\\	
\noindent
Fig. 7 illustrates the difference in variance due to the three detection schemes. As the coherent signal is observed in a thermal bath of Unruh particles, the ideal homodyne scheme observes thermal noise above the shot-noise. As the balanced homodyne detection scheme's phase is different to the phase of the coherent signal, the modal structure that is analysed in the Rindler frame is different to that of the ideal detection scheme. As a result, balanced-homodyne observes a thermal noise which is different to that of ideal. Lastly, the self homodyne sweeps between different phase in the signaller's frame. As a result, the observer sweeps between different modal structures and hence we see a pseudo classical squeezing effect.
\\
\\
We now focus on the observed amplitude. We generally find that the three communication protocols coincides with each other. In Fig. 8, we compare the three detection methods in an interesting regime; when the Minkowski signal is centred around the horizon.	It is noted that the differences between the three communication protocols are due to low frequency contributions in the Rindler frame. This is because the Unruh and Rindler modes are approximately equal to each other in the high frequency limit (i.e. $\hat{c}_{\omega} \approx \hat{a}_{\omega}$ for $\omega/a \gg 1$).	
\begin{figure}[h!]
\includegraphics[width=0.45\textwidth]{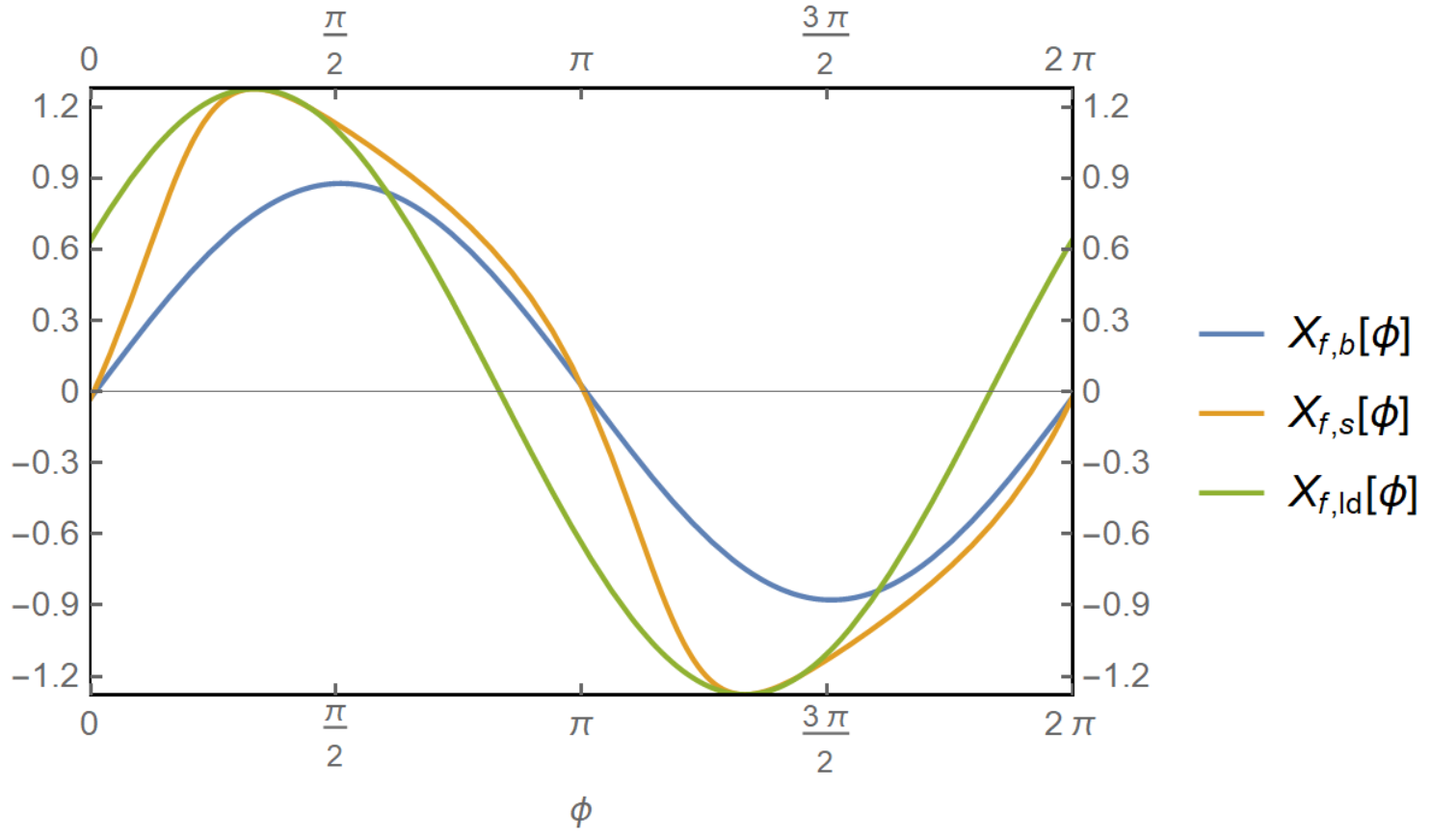}
		\caption{A plot which compares the balanced and self homodyne detection scheme to the idealised homodyne detection scheme. We have utilized the following settings: $a=1,\; V_0=1, \; k_0=1, \; \sigma=0.2, \; |\beta_f|=1, \; \psi= \pi/3,  \; \omega_{min}=10^{-3}$.}	
		\label{fig: Unitary}
	\end{figure}
\\
\noindent
%\\
%\\
%We found that the distortion of the quantum information is related to the local phase of the wave-packet mode at the horizon, and was minimized when the local phase at the horizon was $0$. Some authors [Reference] argue that a phase can only be measured relative to a reference signal. As a result, there are arguments in condensed matter physics that the quantum field should be invariant under a global phase shift [reference]. Our results demonstrate that there are observable difference due to a global phase shift and illustrate the reality of local phase; it is an observable. 
%We demonstrated in the previous section that the noise was suppressed if we set $\omega_{min}$ large enough. We show in Fig. 7 that when $\omega_{min}$ is larger than 1, the three detection schemes coincide with each other. This demonstrates that, the homodyne communication schemes works in the way we expect. However, we notice that the amplitude is amplified to larger than $1$ (we expect an amplitude of 1 at $V_0=0$; half of the signal to be received by the right Rindler observer), demonstrating some minor amplification effects.
We find that there is an amplitude loss and phase shift for balanced-homodyne detection method. These distortion can be traced back to the effects discussed in section VA3. For self-homodyne detection, the phase and amplitude at which the maximum quadrature amplitude measurements occur coincide with the ideal-homodyne detection scheme. On the other hand, the wave-form is largely distorted due to the effects discussed in section VA3. We explicitly observe the effect of phase in-correspondence between the two frames and the effect of phase on the modal shape. One interesting observation is that an orthogonal phase in one frame is not in another frame. We found that this effects emerged due to the phase-information that was lost at the horizon. 
% The amplitude loss can be traced back to the fact that the signal and the local oscillator do not fully overlap in the new reference frame. 
%The phase shift is caused by the fact that the relative phase difference between the two modes are not sustained under change of reference frame. In the self-homodyne, these effect take place every time $\phi$ is changed, distorting the sin waveform. On the other hand, neither of these effects occur when the reference signal is in-phase with the coherent signal. As a result, the self-homodyne coincides with the ideal-homodyne in these two specific cases. 
\\
\\
The presence of horizon leads to tracing out some of the signal that is observed by the observer. The phase information lost through this process leads to information distortion. We analyse another situation to look at other interesting effects.
\subsubsection{Rindler to Delayed Rindler}
In this section we consider sending a Gaussian wave-packet mode signal from the right Rindler frame to the left delayed Rindler frame.  The wave-packet mode is defined as follows:
	\begin{equation}
	\begin{aligned}
	g(\omega;\omega_0,\delta,v_0)& =B\sqrt{\omega}(\frac{1}{2\pi \delta^2})^{1/4} \exp[-\frac{(\omega-\omega_0)^2}{4\delta^2}-i\omega v_0] 
	\end{aligned}
	\end{equation}
where B is the normalization constant. 
\\
\\
The two-mode squeezing between high frequency Rindler and Unruh modes are very small. In this regime, a Gaussian Rindler mode can be approximately transformed into a Minkowski Gaussian wave-packet mode. The results in this regime will be very similar to that obtained in the previous subsection. Fig. 9 looks into the variance of the signal. As the signal is created in a squeezed vacuum, the phase at the horizon does not have a one-to-one correspondence. Due to this we observe a phase distortion effect on top of the noise effect seen in Fig. 7. This effect became negligible as we set $\omega_0$ to be sufficiently large (i.e. when the signal is effectively created in Minkowski vacuum).
\begin{figure}[h!]
\includegraphics[width=0.45\textwidth]{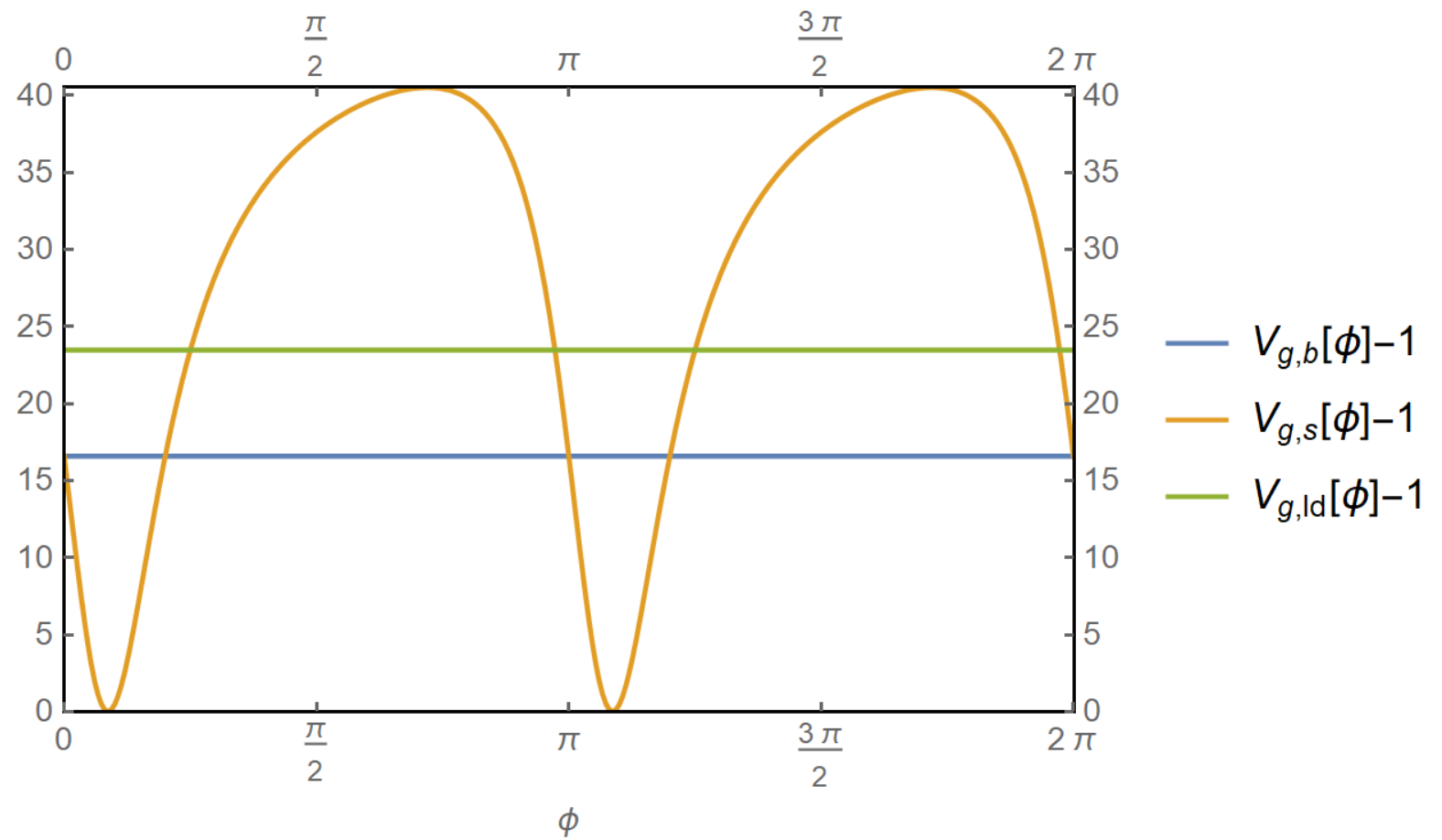}
		\caption{A plot which compares the balanced and self homodyne detection scheme to the idealised homodyne detection scheme. We have utilized the following settings: $a=1,\; v_0=2.5, \; \omega_0=0.5, \; \sigma=0.2, \; |\beta_g|=1, \; \psi= \pi/4, \; \omega_{min}=10^{-3}$.}	
		\label{fig: Unitary}
	\end{figure}
\\
\\
We now move onto looking at the quadrature amplitude. In Fig. 10, we look into the case where we set $\omega_0=0.5$, and $\omega_{min}'=1$. In the Minkowski case, when the low frequency cut-off was set to be sufficiently large, the detection schemes coincided with each other. 
%As a result we are interested in analysing a regime where $\omega_0$ is small enough such that the results are different to that obtained within the previous section, but large enough such that we can set $\omega_{min}=1$. 
We find that the homodyne communication protocols are disturbed even when the low frequency cut-off was set to be sufficiently large. Further analysis shows that this distortion persists even when the the signal is well-localized away from the horizon. We conclude that a signal created in a different two-mode squeezed vacuum leads to distorted QI. We also observe effects similar to that found in the previous subsection if we set $\omega'_{min}$ to be of a smaller value.
\begin{figure}[h!]
\includegraphics[width=0.45\textwidth]{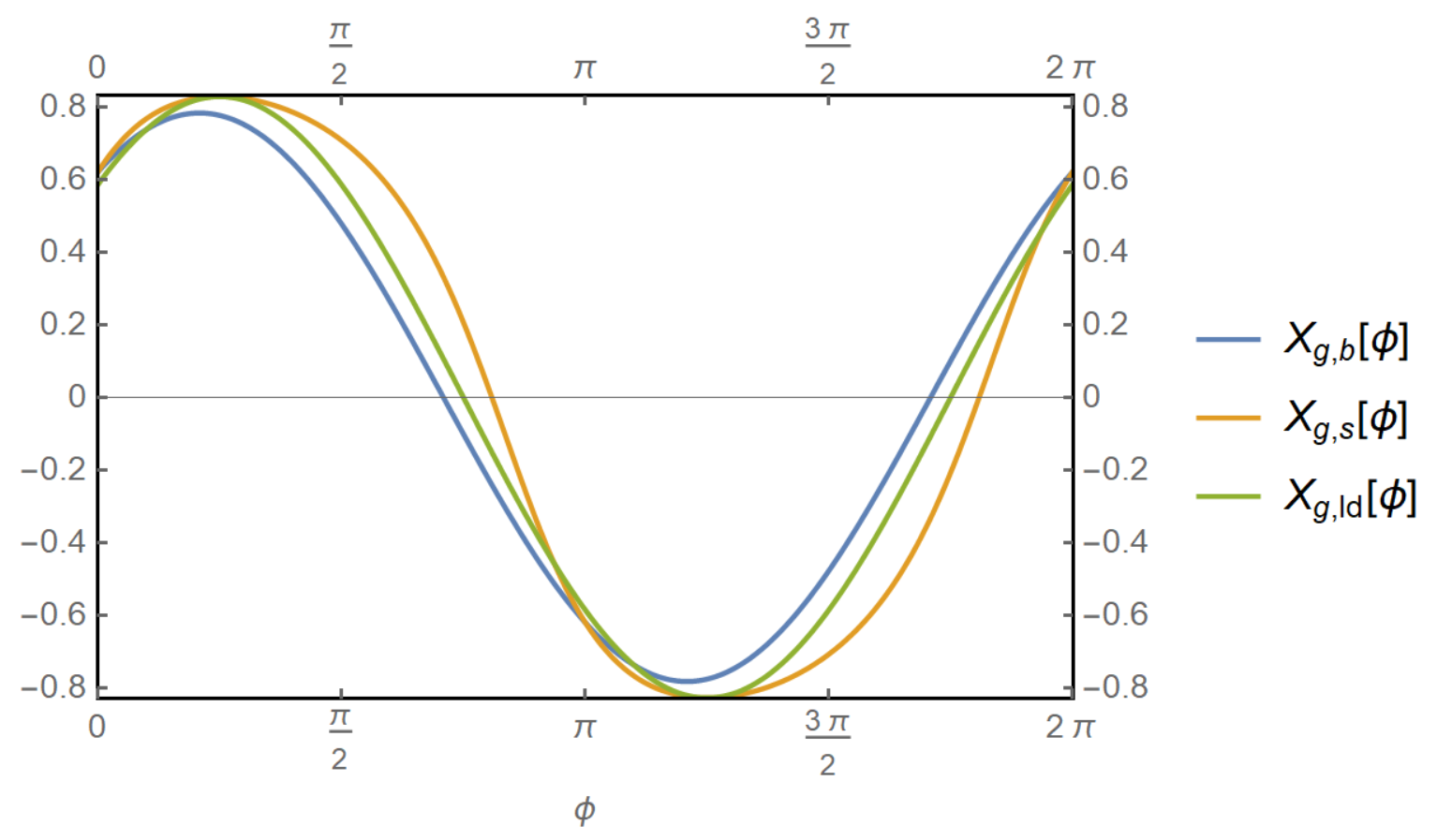}
\caption{A plot which compares the balanced and self homodyne detection scheme to the ideal homodyne detection scheme. We have utilized the following settings: $a=1,\; v_0=2.5, \; \omega_0=0.5, \; \sigma=0.2, |\beta_g|=1, \psi= \pi/4, \; \omega_{min}=1$. }
\label{fig: Unitary}
\end{figure}
\\
\\
This paper introduced the universal transformation of displacement operators. Utilizing this universal transformation, we developed an ideal homodyne detection scheme which allows homodyne communication between different reference frames. We highlighted the issue of negative frequency modes in coherent communication between different frames. We explictly demonstrated the issue by comparing ideal homodyne detection scheme with traditional homodyne detection schemes in two specific scenarios. 
\section{Conclusion}
This paper has studied effects on displacement due to the change of relativistic reference frame. These effects include the distortion of phase information, modal structure and amplitude of the signal. We then demonstrated their effects on traditional homodyne communication via considering coherent signal communication. We highlighted the effect of the horizon in distorting the signal through the analysis of homodyne communication between Minkowski to Rindler. The horizon traces out parts of the signal and the phase information that is lost at the horizon had a significant impact on the homodyne communication. We then highlight the effect of the negative frequency contribution through the analysis of homodyne communication between Rindler to delayed Rindler. As the signal is created in a two-mode squeezed vacuum, the signal naturally carries negative Minkowski frequency modes. The negative frequency modes naturally appeared when the signal was observed in a different non-inertial reference frame.
\\
\\
We utilized the universal transformation of displacement operators to overcome these issues. We developed a homodyne detection technique which is robust to these effects, called the ideal homodyne detection scheme. This detection scheme has the cost that the signaller must know the reference frame they are sending the signal to. The signaller must then accompany the correct coherent signal with the signal so that the observer can extract the full quantum information of the signal. In some scenarios, this may be impractical due to the complexity of the modal structure and/or the need to have access to a space-time region that is space-like separated from the signaller. Nevertheless, this technique is useful in determining the information that is lost throughout the communication to a observer which is in a different non-inertial reference frame.
\\
\\
A straightforward application of ideal homodyne is that it can be utilized to understand the QI of interactions in the Rindler frame. An interesting topic to look into would be the uniformly accelerated mirror. This is a unitarity problem which was raised by Davies and Fulling in 1977 \cite{Davies1977}. The reason why this problem could not be fully resolved can be understood to be due to the incompatibility of the measurement techniques \cite{Davies1977, Birrell1982, Grove1986, Frolov1979, Frolov1980, Frolov1999, Obadia2001, Obadia2003a, Obadia2003b, Svidzinsky2018,Daiqin2017a}. For example, these papers looked at correlation functions \cite{Obadia2003a}, localised statistics in the inertial frame \cite{Daiqin2017a} and excitation of atoms \cite{Svidzinsky2018}. We believe that ideal homodyne would help overcome this issue.
\\
\\
We also highlight that the ideal homodyne tomography can be applied to any reference frame, not just accelerated frames. These include observers in different curvature of space-time \cite{Superposition, Bose2017, Zych2018, Liao2017, Ren2017, Diamanti2017}, various non-inertial trajectories \cite{Hotta2015, Walker1985, Carlitz1987, Wilczek1992, Nicolaevici2003}, and time-varying interactions \cite{Riek2015, Riek2017, Moskalenko2015} including Unruh-DeWitt detectors \cite{Salton2015, Kerstjens2015, Ng2016, Ng2018, Koga2018, Martinez2016,Zhou2018, Martinez2013}. The application of our techniques to such examples would be an interesting future research direction.
\section{Acknowledgement}
We acknowledge support from the Australian
Research Council Centre of Excellence for Quantum
Computation and Communication Technology (Project
No. CE170100012). We also thank Daiqin Su and Robert Mann for useful discussions.

\clearpage
\appendix
\section{Minkowski, Unruh and Rindler modes}
In this paper we consider a massless scalar bosonic field $\hat{\Phi}$ in (1+1)-dimensional Minkowski space-time. Details on the quantisation method and the definition of the single frequency annihlation/creation opeartors can be found in \cite{Unruh1976, Takagi1986, Crispino2008, Fulling1973}. For simplicity, we only consider the left moving modes in this paper. The single frequency Minkowski annihilation (creation) operator is defined as $\hat{e}_k$ ($\hat{e}_k^{\dag}$). The creation and annihilation operators satisfy the bosonic commutation relations: 
\begin{equation}
[\hat{e}_k,\hat{e}_{k'}^{\dag}]=\delta(k-k')
\end{equation}
with all other combination equal to zero.
\\
\\
It is useful to introduce what is known as the single frequency Unruh operators, $\hat{c}_\omega$ and $\hat{d}_\omega$. The Unruh operators are related to the Minkowski operator in the following way \cite{Daiqin2017a, Crispino2008, Birrell1982}:
\begin{align}
&\hat{e}_k=\int d\omega \; A_{k\omega}\hat{c}_{\omega}+B_{k\omega}\hat{d}_{\omega}
\end{align}
	where,
	\begin{equation}
	\begin{aligned}
	A_{k\omega}&=\frac{i\sqrt{2\sinh[\pi\omega/a]}}{2\pi\sqrt{\omega k}}\Gamma[1-i\omega/a]\left(\frac{k}{a}\right)^{i\omega /a} =B_{k\omega}^*
	\end{aligned}
	\end{equation}
Where $\Gamma(x)$ is the gamma function. The expression for $A_{k\omega}$ was obtained by setting the bounds of $e^{i \phi}$ to $-\pi\leq \phi < \pi$. Thus, this convention will be carried out throughout this paper. 
 By inverting the transformation, we find that:
	\begin{equation}
	\begin{aligned}
	\hat{c}_{\omega}&=\int \mathrm{d}k \; A_{k\omega}^* \hat{e}_{k} 
	\\
	\hat{d}_{\omega}&=\int \mathrm{d}k \; B_{k\omega}^* \hat{e}_{k} 
	\end{aligned}
	\end{equation}
	It is easy to show that these function satisfy the following relation:
	\begin{equation}
\begin{aligned}
	& \int_0^{\infty} \mathrm{d}k A_{k\omega}A_{k\omega'}^*=\delta(\omega-\omega')
\\
&      \int_0^{\infty} \mathrm{d}k A_{k\omega}A_{k\omega'}=0
\end{aligned}
	\end{equation}
By utilizing equations (A1, A4, A5), we demonstrate that the Unruh operators must also satisfy the bosonic commutation relations:
\begin{equation}
[\hat{c}_\omega,\hat{c}_{\omega'}^{\dag}]=\delta(\omega-\omega'), \; [\hat{d}_\omega,\hat{d}_{\omega'}^{\dag}]=\delta(\omega-\omega')
\end{equation}
with all other combination equal to zero.
\\
\\
We now introduce a non-inertial observer, Rob, that is accelerated to the right with acceleration $a$. The single frequency modes observed by Rob are referred to as the single frequency right Rindler modes. The corresponding annihilation (creation) operators are denoted as $\hat{a}_{\omega}$ ($\hat{a}_{\omega}^{\dag}$). Likewise, we introduce an observer accelerated to the left, Anti-Rob, and the corresponding single frequency left Rindler annihlation (creation) operators are denoted as $\hat{b}_{\omega}$ ($\hat{b}_{\omega}^{\dag}$). 
\\
\\
The Rindler modes are related to the Unruh modes via a two mode squeezing operation:
	\begin{equation}
	\begin{aligned}
	&\hat{a}_{\omega}= \cosh(r_{\omega}) \hat{c}_{\omega} + \sinh(r_{\omega}) \hat{d}_{\omega}^{\dag} \\
	&\hat{b}_{\omega} 
	= \cosh(r_{\omega}) \hat{d}_{\omega} + \sinh(r_{\omega}) \hat{c}_{\omega}^{\dag} 
	\end{aligned}
	\end{equation}
Where $r_{\omega}\equiv \tanh^{-1}[\exp(-\pi \omega /a)]$ and $a$ is the acceleration of the observer. We can show from equation (A6) and (A7) that the Rindler modes must also satisfy the bosonic commutation relations:
\begin{equation}
[\hat{a}_\omega,\hat{a}_{\omega'}^{\dag}]=\delta(\omega-\omega'), \; [\hat{b}_\omega,\hat{b}_{\omega'}^{\dag}]=\delta(\omega-\omega')
\end{equation}
By inverting equation (A7), we obtain the following equations:
	\begin{equation}
	\begin{aligned}
	&\hat{c}_{\omega}= \cosh(r_{\omega}) \hat{a}_{\omega}- \sinh(r_{\omega}) \hat{b}_{\omega}^{\dag} \\
	&\hat{d}_{\omega} 
	= \cosh(r_{\omega}) \hat{b}_{\omega}- \sinh(r_{\omega}) \hat{a}_{\omega}^{\dag} 
	\end{aligned}
	\end{equation}
Equations (A2), (A4), (A7) and (A9) will form a foundation for the transformation between basis sets. It is noted that we have utilized a different notation to denote the Minkowski, Unruh and Rindler operators to other authors.
\section{Minkowski delayed Rindler modes}
We introduce the Minkowski unitary time evolution operator as follows:
\begin{equation}
\hat{U}=e^{-i \hat{H} t}, \hat{H}\equiv \int \mathrm{d}k\; \hat{e}_k^{\dag} \hat{e}_k k
\end{equation}
We find that the single frequency Minkowski operators evolve under the Heisenberg picture in the following way:
\begin{equation}
\begin{aligned}
\hat{e}_k (t) &\equiv \hat{U}^{\dag}\hat{e}_k (0)\hat{U}
\\& = \hat{e}_k (0) e^{-i k t}
\end{aligned}
\end{equation}
For simplicity, when the time variable is missing, it is assumed that t=0. By utilizing this result, we find that the Unruh modes evolve under the unitary in the following way:
\begin{equation}
\begin{aligned}
\hat{c}_{\omega}(t) &\equiv \hat{U}^{\dag}\hat{c}_{\omega} \hat{U}= \int \mathrm{d}k A_{k\omega}^*e^{-ikt} \hat{e}_k 
\\
\hat{d}_{\omega}(t) &=  \int \mathrm{d}k A_{k\omega}e^{-ikt} \hat{e}_k 
\end{aligned}
\end{equation}
As the Unruh modes forms a complete basis, we know that the Minkowski evolved Unruh modes can be decomposed in the following way:
\begin{equation}
\begin{aligned}
\hat{c}_{\omega}(t) & = \int \mathrm{d}\omega'\; A_{\omega,\omega'} (t) \hat{c}_{\omega'}+B_{\omega,\omega'}(t) \hat{d}_{\omega'}
\\
\hat{d}_{\omega}(t) & = \int \mathrm{d}\omega'\; C_{\omega,\omega'}(t) \hat{c}_{\omega'}+D_{\omega,\omega'}(t) \hat{d}_{\omega'}
\end{aligned}
\end{equation}
The Rindler modes evolve under the unitary in the following way:
\begin{equation}
\begin{aligned}
\hat{a}_{\omega}(t) &=  \int \mathrm{d}k \; \cosh(r_{\omega})A_{k\omega}e^{-ikt} \hat{e}_k+\sinh(r_{\omega})A_{k\omega}e^{ikt} \hat{e}_k ^{\dag}
\\
\hat{b}_{\omega}(t) &=  \int \mathrm{d}k \; \cosh(r_{\omega}) A_{k\omega}^*e^{-ikt} \hat{e}_k+ \sinh(r_{\omega})A_{k\omega}^* e^{ikt} \hat{e}_k ^{\dag}
\end{aligned}
\end{equation}
As the Rindler modes also forms a complete basis, the Minkowski evolved Rindler modes can be decomposed in the following way:
\begin{widetext}
\begin{equation}
\begin{aligned}
\hat{a}_{\omega}(t) &= \int \mathrm{d}\omega'\; {\alpha}^a_{\omega,\omega'} (t) \hat{a}_{\omega'}+{\beta}^a_{\omega,\omega'}(t) \hat{a}_{\omega'}^{\dag}+{\gamma}^a_{\omega,\omega'}(t) \hat{b}_{\omega'}+{\delta}^a_{\omega,\omega'}(t) \hat{b}_{\omega'}^{\dag}
\\
\hat{b}_{\omega}(t) &= \int \mathrm{d}\omega'\; {\alpha}^b_{\omega,\omega'}(t) \hat{a}_{\omega'}+{\beta}^b_{\omega,\omega'}(t) \hat{a}_{\omega'}^{\dag}+{\gamma}^b_{\omega,\omega'}(t) \hat{b}_{\omega'}+{\delta}^b_{\omega,\omega'}(t) \hat{b}_{\omega'}^{\dag}
\end{aligned}
\end{equation}
\begin{equation}
\begin{aligned}
\hat{a}_{\omega} &= \int \mathrm{d}\omega'\; {\alpha}^a_{\omega',\omega}{}^* (t) \hat{a}_{\omega'}(t)-{\beta}^a_{\omega',\omega}(t) \hat{a}_{\omega'}(t)^{\dag}+{\alpha}^b_{\omega',\omega}{}^*(t) \hat{b}_{\omega'}(t)-{\beta}^b_{\omega',\omega}(t) \hat{b}_{\omega'}(t)^{\dag}
\\
\hat{b}_{\omega} &= \int \mathrm{d}\omega'\; {\gamma}^a_{\omega,\omega'}{}^*(t) \hat{a}_{\omega'}(t)-{\delta}^a_{\omega,\omega'}(t) \hat{a}_{\omega'}(t)^{\dag}+{\gamma}^b_{\omega,\omega'}{}^*(t) \hat{b}_{\omega'}(t)-{\delta}^b_{\omega,\omega'}(t) \hat{b}_{\omega'}(t)^{\dag}
\end{aligned}
\end{equation}
\end{widetext}
The explicit expressions of the Bogoliubov coefficients in equation (B4) and (B6) are calculated in the following section. We introduce Anti-Rob' and Rob' in Fig. 2. These observers can be considered as the observer who observe the single frequency Minkowski delayed Rindler modes; equation (B6).
\begin{widetext}
\section{Bogoliubov Coefficients for Minkowski Delayed Unruh Modes}
The Bogoliubov coefficients can be calculated by taking the commutator between the two operators. 
\begin{equation}
\begin{aligned}
{} [\hat{c}_{\omega}(t),\hat{c}_{\omega'}^{\dag}] & = [\int \mathrm{d}\alpha\; A_{\omega,\alpha} \hat{c}_{\alpha}+B_{\omega,\alpha} \hat{d}_{\alpha},\hat{c}_{\omega'}^{\dag} ]=A_{\omega,\omega'}
\end{aligned}
\end{equation}
Likewise, we find the following:
\begin{equation}
\begin{aligned}
B_{\omega,\omega'} & =[\hat{c}_{\omega}(t),\hat{d}_{\omega'}^{\dag}], \; C_{\omega,\omega'} =[\hat{d}_{\omega}(t),\hat{c}_{\omega'}^{\dag}], \; D_{\omega,\omega'} =[\hat{d}_{\omega}(t),\hat{d}_{\omega'}^{\dag}]
\end{aligned}
\end{equation}
Thus, we can calculate the Bogoliubov transformation coefficients by utilizing equation (A4) and (B3) and explicitly calculating the commutators given in equation (C1) and (C2). Below, we explicitly calculate $A_{\omega,\omega'}$.
\begin{equation}
\begin{aligned}
{} A_{\omega,\omega'} &= [\int \mathrm{d}k A_{k\omega}^*e^{-ikt} \hat{e}_k ,\int \mathrm{d}k' A_{k'\omega'}\hat{e}_{k'}^{\dag} ]
\\ &=  \int \mathrm{d}k A_{k\omega}^* A_{k\omega'}e^{-ikt} 
\\ &=  \frac{\sqrt{\sinh (\pi \omega /a) \sinh (\pi \omega' /a)}}{2 \pi^2 \sqrt{\omega \omega'}}\Gamma[1+i \omega/a]\Gamma[1 - i \omega' /a] \int \mathrm{d}k \frac{1}{k} (\frac{k}{a})^{-i (\omega/a-\omega'/a)}e^{-ikt} 
\end{aligned}
\end{equation}
For $t>0$, the integral evaluates to the following:
\begin{equation}
\begin{aligned}
\int \mathrm{d}k \; \frac{1}{k} (\frac{k}{a})^{-i (\omega/a-\omega'/a)}e^{-ikt} & = (i at)^{i (\omega-\omega ')/a} \Gamma[-i (\omega-\omega')/a]
\\ &= e^{- \pi (\omega-\omega') /(2a)}(a|t|)^{i (\omega-\omega ')/a} \Gamma[-i (\omega-\omega')/a]
\end{aligned}
\end{equation}
Where the second line utilized the definition that $i=e^{i \pi/2}$. When we induce a time-delay the time is negative, and hence $t=-|t|$. In this case, we utilize the fact that $-i= e^{-i\pi/2}$ and find that:
\begin{equation}
\begin{aligned}
\int \mathrm{d}k \; \frac{1}{k} (\frac{k}{a})^{-i (\omega/a-\omega'/a)}e^{-ikt} & = e^{\pi (\omega-\omega') /(2a)} (a|t|)^{i (\omega-\omega ')/a} \Gamma[-i (\omega-\omega')/a]
\end{aligned}
\end{equation}
%Utilizing the fact that $\Gamma[1+i|x|]\Gamma[1-i|x|]=\frac{\pi |x|}{\sinh(\pi |x|)}$ and $\Gamma[1+x]=x\Gamma[x]$, we find that $A_{\omega,\omega'}$ simplifies to the following value:
\begin{equation}
\begin{aligned}
A_{\omega,\omega'}(\pm |t|) &= Z_{\omega}^*Z_{\omega'}e^{\mp \frac{\pi}{2a}(\omega-\omega')} (a|t|)^{i(\omega-\omega')/a}\Gamma(-i(\omega-\omega'))
\end{aligned}
\end{equation}
Where we have defined the following:
\begin{equation}
\begin{aligned}
Z_{\omega}\equiv \frac{i \sqrt{2 \sinh{\pi \omega/a}}}{2 \pi \sqrt{\omega}} \Gamma(i-i \omega/a)
\end{aligned}
\end{equation}
By following similar steps, we find the following:
\begin{equation}
\begin{aligned}
B_{\omega,\omega'}(\pm |t|) &= Z_{\omega}^*Z_{\omega'}^*e^{\mp \frac{\pi}{2a}(\omega+\omega')} (a|t|)^{i(\omega+\omega')/a}\Gamma(-i(\omega+\omega'))
\\
C_{\omega,\omega'}(\pm |t|) &= Z_{\omega}Z_{\omega'}e^{\pm \frac{\pi}{2a}(\omega+\omega')} (a|t|)^{-i(\omega+\omega')/a}\Gamma(i(\omega+\omega'))
\\
D_{\omega,\omega'}(\pm |t|) &=Z_{\omega}Z_{\omega'}^* e^{\pm \frac{\pi}{2a}(\omega-\omega')} (a|t|)^{-i(\omega-\omega')/a}\Gamma(i(\omega-\omega')) 
\end{aligned}
\end{equation}
We find that these coefficients are related in the following way:
\begin{equation}
\begin{aligned}
A_{\omega,\omega'}(\pm |t|)=D_{\omega,\omega'}(\mp |t|)^*
\\B_{\omega,\omega'}(\pm |t|) = C_{\omega,\omega'}(\mp |t|)^*
\end{aligned}
\end{equation}
\section{Bogoliubov Coefficients for Minkowski Delayed Rindler Modes}
We use the following properties to calculate the Bogoliubov coefficients for the Rindler modes:
	\begin{equation}
	\begin{aligned}
	&\hat{c}_{\omega} (t') = \cosh(r_{\omega}) \hat{a}_{\omega}(t')- \sinh(r_{\omega}) \hat{b}_{\omega}^{\dag} (t') \\
	&\hat{d}_{\omega} (t')
	= \cosh(r_{\omega}) \hat{b}_{\omega}(t')- \sinh(r_{\omega}) \hat{a}_{\omega}^{\dag} (t')
	\end{aligned}
	\end{equation}
	\begin{equation}
	\begin{aligned}
	&\hat{a}_{\omega}(t')= \cosh(r_{\omega}) \hat{c}_{\omega}(t') + \sinh(r_{\omega}) \hat{d}_{\omega}^{\dag}(t') \\
	&\hat{b}_{\omega} (t')
	= \cosh(r_{\omega}) \hat{d}_{\omega}(t') + \sinh(r_{\omega}) \hat{c}_{\omega}^{\dag} (t')
	\end{aligned}
	\end{equation}
%It is noted that this property is satisfied for all $t'$, and hence we can set $t'=0$. 
By utilizing equation (B4) and  (D2), we find the following:
\begin{equation}
\begin{aligned}
	\hat{a}_{\omega}(t)&= \cosh(r_{\omega})\left( \int \mathrm{d}\omega'\; A_{\omega,\omega'} (t) \hat{c}_{\omega'}+B_{\omega,\omega'}(t) \hat{d}_{\omega'} \right)+ \sinh(r_{\omega}) \left(\int \mathrm{d}\omega'\; C_{\omega,\omega'}(t)^* \hat{c}_{\omega'}^{\dag}+D_{\omega,\omega'}(t)^* \hat{d}_{\omega'}^{\dag}\right)
\\
	\hat{b}_{\omega} (t) &= \cosh(r_{\omega}) \left(\int \mathrm{d}\omega'\; C_{\omega,\omega'}(t) \hat{c}_{\omega'}+D_{\omega,\omega'}(t) \hat{d}_{\omega'}\right) + \sinh(r_{\omega}) \left(\int \mathrm{d}\omega'\; A_{\omega,\omega'} (t)^* \hat{c}_{\omega'}^{\dag}+B_{\omega,\omega'}(t)^* \hat{d}_{\omega'}^{\dag}\right)
\end{aligned}
\end{equation}
By utilizing equation (A9), we can compare this equation with equation (B6). We find that the coefficients are equal to the following
\begin{equation}
\begin{aligned}
 \\ {\alpha}^a_{\omega,\omega'}(t)&= \cosh(r_{\omega})\cosh(r_{\omega'})A_{\omega,\omega'} (t)-\sinh(r_{\omega})\sinh(r_{\omega'})A_{\omega,\omega'}(-t) 
\\{\beta}^a_{\omega,\omega'} (t) &= -\cosh(r_{\omega})\sinh(r_{\omega'})B_{\omega,\omega'} (t)+\sinh(r_{\omega})\cosh(r_{\omega'})B_{\omega,\omega'}(-t)
\\{\gamma}^a_{\omega,\omega'}(t) &=\cosh(r_{\omega})\cosh(r_{\omega'})B_{\omega,\omega'} (t)-\sinh(r_{\omega})\sinh(r_{\omega'})B_{\omega,\omega'}(-t) 
\\{\delta}^a_{\omega,\omega'} (t) &=-\cosh(r_{\omega})\sinh(r_{\omega'})A_{\omega,\omega'} (t)+\sinh(r_{\omega})\cosh(r_{\omega'})A_{\omega,\omega'}(-t)
\end{aligned}
\end{equation}
\begin{equation}
\begin{aligned}
 {\alpha}^b_{\omega,\omega'} (t)&= \cosh(r_{\omega})\cosh(r_{\omega'})C_{\omega,\omega'} (t)-\sinh(r_{\omega})\sinh(r_{\omega'})C_{\omega,\omega'}(-t) 
\\{\beta}^b_{\omega,\omega'}(t) &= -\cosh(r_{\omega})\sinh(r_{\omega'})D_{\omega,\omega'} (t)+\sinh(r_{\omega})\cosh(r_{\omega'})D_{\omega,\omega'}(-t)
\\{\gamma}^b_{\omega,\omega'}(t)&=\cosh(r_{\omega})\cosh(r_{\omega'})D_{\omega,\omega'} (t)-\sinh(r_{\omega})\sinh(r_{\omega'})D_{\omega,\omega'}(-t) 
\\{\delta}^b_{\omega,\omega'} (t) &=-\cosh(r_{\omega})\sinh(r_{\omega'})C_{\omega,\omega'} (t)+\sinh(r_{\omega})\cosh(r_{\omega'})C_{\omega,\omega'}(-t)
\end{aligned}
\end{equation}
By utilizing the relation defined in equation  (C9), we find that these coefficients are related in the following way:
\begin{equation}
\begin{aligned}
 \\ {\alpha}^a_{\omega,\omega'}(t)&= {\gamma}^b_{\omega,\omega'}(-t)^*
\\{\gamma}^a_{\omega,\omega'}(t) &={\alpha}^b_{\omega,\omega'}(-t)^*
\\{\beta}^a_{\omega,\omega'} (t) &= {\delta}^b_{\omega,\omega'} (-t)^*
\\{\delta}^a_{\omega,\omega'} (t) &={\beta}^a_{\omega,\omega'} (-t)^*
\end{aligned}
\end{equation}
We simplify ${\alpha}^a_{\omega,\omega'}(t)$ by considering ${\alpha}^a_{\omega,\omega'}(|t|)$ and ${\alpha}^a_{\omega,\omega'}(-|t|)$ separately. We make use of a property of the gamma function, $|\Gamma (1+i x)|^2=\frac{\pi x}{\sinh(\pi x)}$, to simplify the expression.
\begin{equation}
\begin{aligned} 
{\alpha}^a_{\omega,\omega'} (|t|)&=A_{\omega,\omega'} (|t|)( \cosh(r_{\omega})\cosh(r_{\omega'})-\sinh(r_{\omega})\sinh(r_{\omega'})e^{\pi (\omega-\omega')/a} )
 \\ &=A_{\omega,\omega'}(|t|) e^{\pi (\omega- \omega')/2a} \frac{\sqrt{\sinh(\pi \omega' /a)}}{\sqrt{\sinh(\pi \omega /a)}}
 \\ &=i \sqrt{\frac{\omega'}{\omega}} \frac{1}{2 \pi (\omega-\omega')/a} \frac{\Gamma(1+i \omega/a) \Gamma(1-i(\omega-\omega'))}{\Gamma(1+i \omega'/a)}(a|t|)^{i(\omega-\omega')/a}
\\
  {\alpha}^a_{\omega,\omega'} (-|t|) &=A_{\omega,\omega'}(|t|) e^{\pi (\omega- \omega')/2a} \frac{\sqrt{\sinh(\pi \omega /a)}}{\sqrt{\sinh(\pi \omega' /a)}}
\\  &=i \sqrt{\frac{\omega}{\omega'}} \frac{1}{2 \pi (\omega-\omega')/a} \frac{\Gamma(1-i \omega'/a) \Gamma(1-i(\omega-\omega'))}{\Gamma(1-i \omega/a)}(a|t|)^{i(\omega-\omega')/a}
\end{aligned}
\end{equation}
Following similar steps, we find that:
\begin{equation}
\begin{aligned} 
{\beta}^a_{\omega,\omega'} (|t|)&= B_{\omega,\omega'}(|t|)e^{\pi(\omega+\omega')/2a}\frac{\sqrt{\sinh(\pi \omega'/a)}}{\sqrt{\sinh(\pi \omega/a)}}
\\&=- i \sqrt{\frac{\omega'}{\omega}} \frac{1}{2 \pi (\omega+\omega')/a} \frac{\Gamma(1+i \omega/a) \Gamma(1-i(\omega+\omega'))}{\Gamma(1-i \omega'/a)}(a|t|)^{i(\omega+\omega')/a}
\\
  {\beta}^a_{\omega,\omega'} (-|t|)&= B_{\omega,\omega'}(|t|)e^{\pi(\omega+\omega')/2a}\frac{-\sqrt{\sinh(\pi \omega/a)}}{\sqrt{\sinh(\pi \omega'/a)}}
  \\&= i \sqrt{\frac{\omega}{\omega'}} \frac{1}{2 \pi (\omega+\omega')/a} \frac{\Gamma(1+i \omega'/a) \Gamma(1-i(\omega+\omega'))}{\Gamma(1-i \omega/a)}(a|t|)^{i(\omega+\omega')/a}
 \\
 {\gamma}^a_{\omega,\omega'} (|t|)&=0
\\
  {\gamma}^a_{\omega,\omega'} (-|t|)&=B_{\omega,\omega'}(|t|)e^{\pi(\omega+\omega')/2a} \frac{\sinh(\pi(\omega+\omega'))}{\sqrt{\sinh(\pi \omega/a)}\sqrt{\sinh(\pi \omega'/a)}}
  \\&= i \frac{1}{2 \pi\sqrt{\omega \omega'}} \frac{\Gamma(1+i \omega/a)\Gamma(1+i \omega'/a) }{\Gamma(1+i(\omega+\omega')/a)}(a|t|)^{i(\omega+\omega')/a}
 \\
 {\delta}^a_{\omega,\omega'} (|t|)&=0
\\
  {\delta}^a_{\omega,\omega'} (-|t|)&=A_{\omega,\omega'}(|t|)e^{\pi(\omega-\omega')/2a} \frac{\sinh(\pi(\omega+\omega'))}{\sqrt{\sinh(\pi \omega/a)}\sqrt{\sinh(\pi \omega'/a)}}
\\ 
  &=-i \frac{1}{2 \pi\sqrt{\omega \omega'}} \frac{\Gamma(1+i \omega/a)\Gamma(1-i \omega'/a) }{\Gamma(1+i(\omega-\omega')/a)}(a|t|)^{i(\omega-\omega')/a}
\end{aligned}
\end{equation}
Through these calculations, we have derived the Bogoliubov transformation coefficients between Minkowski time-evolved/delayed Rindler/Unruh modes. In our paper, we are particularly interested in $\alpha^{b}_{\omega,\omega'}(|t|)$ and $\beta^{b}_{\omega,\omega'}(|t|)$, so we explicitly write their expressions here:
\begin{equation}
\begin{aligned} 
\alpha^{b}_{\omega,\omega'}(|t|)&=-i \frac{1}{2 \pi \sqrt{\omega \omega'}}\frac{\Gamma(1- i \omega/a) \Gamma(1-i \omega' /a)}{\Gamma(1-i(\omega+\omega')/a)} (a|t|)^{-i (\omega+\omega')/a}
\\
\beta^{b}_{\omega,\omega'}(|t|)=&i \frac{1}{2 \pi \sqrt{\omega \omega'}}\frac{\Gamma(1 - i \omega/a) \Gamma(1 + i \omega' /a)}{\Gamma(1-i(\omega-\omega')/a)} (a|t|)^{-i (\omega+\omega')/a}
\end{aligned}
\end{equation}
\end{widetext}

\end{document}